\documentclass{aastex631}

\begin{document}

%\title{A Polarized Image-based Search for Pulsars Toward the Galactic Bulge}

\title{An Image-Based Search for Pulsar Candidates in the MeerKAT Bulge Survey}

\author[0009-0008-7604-003X]{Dale A. Frail}
\affiliation{National Radio Astronomy Observatory, P.O. Box O, Socorro, NM 87801, USA}

\author[0000-0003-3272-9237]{Emil Polisensky}
\affiliation{U.S.\ Naval Research Laboratory,  4555 Overlook Ave SW,  Washington,  DC 20375,  USA}

\author[0009-0006-5070-6329]{Scott D. Hyman}
\affiliation{Department of Engineering and Physics, Sweet Briar College, Sweet Briar, VA 24595, USA}

\author[0000-0001-7363-6489]{William D. Cotton}
\affiliation{National Radio Astronomy Observatory, 520 Edgemont Road, Charlottesville, VA 22903, USA}

\author[0000-0001-8035-4906]{Namir E. Kassim}
\affiliation{U.S.\ Naval Research Laboratory,  4555 Overlook Ave SW,  Washington,  DC 20375,  USA}

\author[0000-0003-2565-7909]{Michele L. Silverstein}
\altaffiliation{NRC Research Associate}
\affiliation{U.S.\ Naval Research Laboratory,  4555 Overlook Ave SW,  Washington,  DC 20375,  USA}

\author[0000-0002-9409-3214]{Rahul Sengar}
\affiliation{Center for Gravitation, Cosmology, and Astrophysics, Department of Physics, University of Wisconsin-Milwaukee, PO Box 413, Milwaukee, WI 53201, USA}

\author[0000-0001-6295-2881]{David L. Kaplan}
\affiliation{Center for Gravitation, Cosmology, and Astrophysics, Department of Physics, University of Wisconsin-Milwaukee, PO Box 413, Milwaukee, WI 53201, USA}

\author[0000-0001-7722-6145]{Francesca Calore}
\affiliation{Laboratoire d’Annecy-le-Vieux de Physique Théorique, CNRS, F-74000 Annecy, France}

\author[0000-0003-4962-145X]{Joanna Berteaud}
\affiliation{University of Maryland, Department of Astronomy, College Park, MD 20742, USA}
\affiliation{NASA Goddard Space Flight Center, Code 662, Greenbelt, MD 20771, USA}

\author[0000-0003-0724-2742]{Ma\"ica Clavel}
\affiliation{Université Grenoble Alpes, CNRS, IPAG, F-38000 Grenoble, France}

\author[0000-0002-2822-1919]{Marisa Geyer}
\affiliation{High Energy Physics, Cosmology and  Astrophysics Theory (HEPCAT) Group, Department of Mathematics and Applied Mathematics, University of Cape Town,  Rondebosch 7701, South Africa}
\affiliation{South African Radio Astronomy Observatory, 2 Fir Street, Black River Park, Observatory 7925, South Africa}

\author[0000-0001-5205-8501]{Samuel Legodi}
\affiliation{South African Radio Astronomy Observatory, 2 Fir Street, Black River Park, Observatory 7925, South Africa}

\author[0000-0002-1966-5729]{Vasaant Krishnan}
\affiliation{South African Radio Astronomy Observatory, 2 Fir Street, Black River Park, Observatory 7925, South Africa}

\author[0000-0002-1691-0215]{Sarah Buchner}
\affiliation{South African Radio Astronomy Observatory, 2 Fir Street, Black River Park, Observatory 7925, South Africa}

\author[0000-0002-1873-3718]{Fernando Camilo}
\affiliation{South African Radio Astronomy Observatory, 2 Fir Street, Black River Park, Observatory 7925, South Africa}

\correspondingauthor{Dale A. Frail}
\email{dfrail@nrao.edu}

\begin{abstract}
We report on the results of an image-based search for pulsar candidates toward the Galactic bulge. We used mosaic images from the MeerKAT radio telescope, that were taken as part of a 173 deg$^2$ survey of the bulge and Galactic center of our Galaxy at L band (856-1712 MHz) in all four Stokes  I, Q, U and V. The image root-mean-square noise levels of 12-17 $\mu$Jy ba$^{-1}$ represent a significant increase in sensitivity over past image-based pulsar searches. Our primary search criterion was circular polarization, but we used other criteria including linear polarization, in-band spectral index, compactness, variability and multi-wavelength counterparts to select pulsar candidates.  We first demonstrate the efficacy of this technique by searching for polarized emission from known pulsars, and comparing our results with measurements from the literature. Our search resulted in a sample of 75 polarized sources. Bright stars or young stellar objects were associated with 28 of these sources, including a small sample of highly polarized dwarf stars with pulsar-like steep spectra. Comparing the properties of this sample with the known pulsars, we identified 30 compelling candidates for pulsation follow-up, including two sources with both strong circular and linear polarization. The remaining 17 sources are either pulsars or stars, but we cannot rule out an extragalactic origin or image artifacts among the brighter, flat spectrum objects.
\end{abstract}



\section{Introduction} \label{sec:intro}

With the advent of a new generation of sensitive wide-field radio interferometers, there has been a renewed interest in imaging the Galactic center (GC) and the bulge of our Galaxy \citep{2022ApJ...925..165H,2022MNRAS.516.5972W,2023arXiv231207275G}. Early pioneering efforts revealed a significant population of compact radio sources \citep{2000AJ....119..207L, 2004AJ....128.1646N, 2008ApJS..174..481L}, while time-domain monitoring led to the discovery of new classes of transients and variables objects \citep[e.g.,][]{1992Sci...255.1538Z,2005Natur.434...50H,2021ApJ...920...45W,2022ApJ...927L...6Z}.

Another important application of image-based radio surveys is the identification of pulsar candidates \citep[e.g.,][]{2000ApJ...529..859K,2018MNRAS.474.5008D}. It has long been argued that the enhanced stellar densities in the GC and Galactic bulge could lead to a significant number of pulsars, and in particular millisecond pulsars \citep{2006ApJS..165..173M,2015ApJ...805..172M,2021PhRvD.104d3007B}. There is an on-going debate on whether the diffuse gamma-ray emission from the Galactic bulge, the so-called gamma-ray excess identified by the {\it Fermi} Large Area Gamma-Ray Telescope, originates from a population of thousands (or more) millisecond pulsars (MSPs) or is the first, non-gravitational, signature of particles dark matter annihilation \citep[e.g.,][]{Calore:2021jvg, Song:2024iup}, see \cite{2020ARNPS..70..455M} for a review. Identifying pulsations from this putative MSP population would go a long way in resolving this controversy, but the existing census of pulsars is inadequate \citep{2016ApJ...827..143C}. Traditional pulsation searches suffer from a number of special challenges in this direction \citep{2016ApJ...827..143C} but they can be effective when used together with image-based methods \citep{2018MNRAS.475..942F, 2024ApJ...969...30M,2023arXiv231114880W,2024arXiv240102498A}, also combined with multi-wavelength cross-correlations~\citep{Berteaud:2023xjp}. 

The initial image-based pulsar searches in the Galactic bulge and GC relied on only two pulsar properties: their compactness and steep spectral indices \citep{2017MNRAS.468.2526B,2019ApJ...876...20H}. Variability has been re-discovered as a sufficient but not necessary condition for identifying promising pulsar candidates \citep{2022MNRAS.516.5972W, 2023MNRAS.525L..76H,2023PASA...40....3S}, but polarization has proven to be an even more powerful search criterion. Pulsars appear to be unique among compact radio sources for having both steep spectra and high degrees of linear and/or circular polarization \citep{2018MNRAS.474.4629J,2021MNRAS.504..228S}. Identifying pulsar candidates by a high degree of polarization in the image plane was first used to identify the young, energetic pulsar PSR\,B1951+32 in the supernova remnant CTB\,80 \citep{1987ApJ...319L.103S}, and then again to identify the luminous MSP PSR\,J0218+4232 \citep{1995ApJ...455L..55N}. Recently there has been a resurgence in imaging surveys with an emphasis on polarization \citep{2018MNRAS.478.2835L,2022A&A...661A..87S,2022ApJ...930...38W,2023PASA...40...34D,2023A&A...670A.124C}. There has been some initial success at identifying promising pulsar candidates in the GC and bulge using polarization as an additional search criterion \citep{2021MNRAS.507.3888H,2022ApJ...930...38W,2022MNRAS.516.5972W}. 

In this paper we build upon these efforts by using a new sensitive survey of the Galactic bulge and GC region in full Stokes in a search for pulsar candidates. The paper is organized in the following sections: 
we present the data analyzed and data reduction pipeline in section~\ref{sec:obs} and~\ref{sec:reduce}
respectively. In section~\ref{sec:catalog}, we illustrate the method followed to build our main source 
catalog, including a discussion of survey properties and known data issues.
In section~\ref{sec:search}, we put forward our selection strategy for pulsar candidates, mainly based on polarization characteristics.
In section~\ref{sec:count}, we comment on other important selection criteria. 
The main results are presented in section~\ref{sec:results}, where we present the pulsar candidate list and a detailed discussion about the non-detection of some of these in previous pulsation-searches campaigns. 
We conclude in section~\ref{sec:discuss}.

\section{Observations}\label{sec:obs}

For this work we have used an archival South African Radio Astronomy Observatory (SARAO) Legacy Survey of the Galactic Center Region (Code SSV-20180505-FC-01), taken with the 64-element MeerKAT Radio Telescope array \citep{2016mks..confE...1J,2018ApJ...856..180C,2020ApJ...888...61M}. A  vertical strip at approximately $\vert{b}\vert< 20^\circ$ and at least $\vert{l}\vert< 1.5^\circ$ (widening to $\pm{4.5}^\circ$ at lower latitudes) was observed toward the GC, using the L-band receivers (856-1712 MHz). A total of 315 individual pointings were made during 42 sessions from 2019 Dec. 26 to 2020 Aug. 5. A single session was typically 8-10 hours in duration, during which 7 or 8 pointings were observed, each receiving approximately one hour in total integration time. Pointings were widely spaced in hour angle to maximize spatial frequency ($uv$) coverage. 
%https://skaafrica.atlassian.net/wiki/spaces/ESDKB/pages/1382154297/Science+with+MeerKAT#The-Galactic-Center-region

The MeerKAT correlator was configured to produce 4096 channels across the full bandwidth in 8-s sampling in all four combinations (XX, YY, XY, and 
YX) of the orthogonal linearly polarized feeds. PKS B1934$-$638 was used as the photometric and bandpass calibrator, 3C\,286 as the polarization calibrator, and a nearby astrometric calibrator was used for each session. For a similar observational setup see \cite{2022ApJ...925..165H}, or consult the MeerKAT Knowledge Base\footnote{\url{https://skaafrica.atlassian.net/wiki/spaces/ESDKB/overview?homepageId=41025669}} for current best practices.

\section{Data Reduction}\label{sec:reduce}

All calibration, imaging and mosaicing was carried out in the Obit software package\footnote{\url{http://www.cv.nrao.edu/~bcotton/Obit.html}} following standard procedures. We summarize this process here but a full description is given in the recently released SARAO MeerKAT 1.3 GHz Galactic Plane Survey \citep{2023arXiv231207275G}.

Calibration and editing followed procedures established by \cite{2020ApJ...888...61M}, in which instrumental data errors and radio interference were identified and flagged. The data were then calibrated for phase, amplitude, bandpass, and group delays as described by \cite{2022A&A...657A..56K}. Polarization calibration followed the same procedure outlined in \cite{2023arXiv231207275G}, using 3C\,286 as a polarized calibrator, and PKS B1934$-$638 as an unpolarized source.

\begin{figure}[htb!]
\includegraphics[width=10cm]{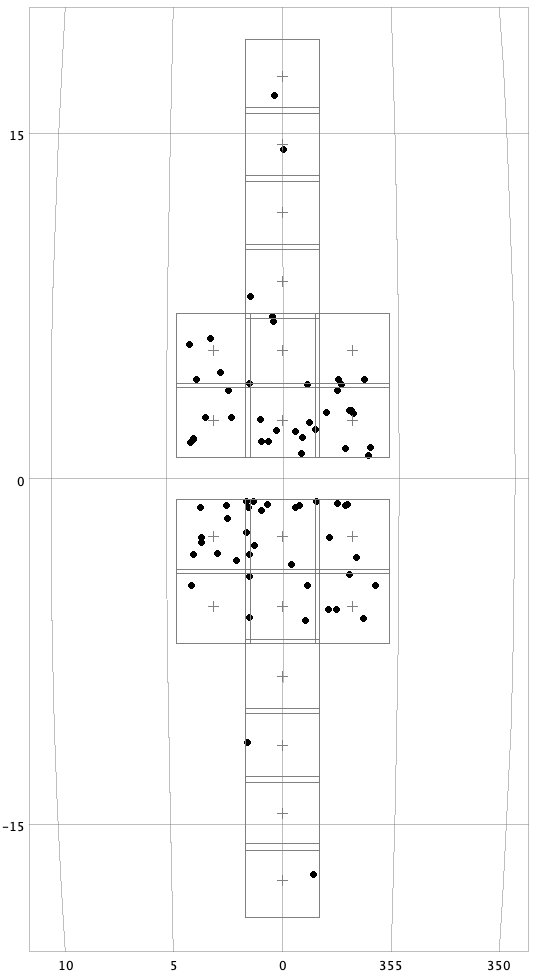}\label{fig:survey}
\centering
\caption{Galactic distribution of the 20 mosaic pointings used in this survey. Each square is 3.125$^\circ$ × 3.125$^\circ$ in size. The full survey symmetric about the GC and covers a vertical strip $\pm{20}^\circ$ in latitude and at least $\pm{1.5}^\circ$ in longitude, but widens to $\pm{4.5}^\circ$ at lower latitudes. The positions of the known pulsars that lie within the survey area are indicated by filled circles.}
\end{figure}

Once calibrated the individual pointings in all four Stokes I, Q, U and V were imaged and deconvolved within Obit. These individual pointings were formed into 20 partially overlapping linear mosaics each 3.125$^\circ$×3.125$^\circ$ in size using the optimal weighting scheme described in \cite{2023arXiv231207275G}. Each mosaic was also imaged in 14 subbands, with bandwidths ranging from 43 to 74~MHz. However, the middle two subbands were severely affected by radio frequency interference (RFI) and had to be discarded. As a result, there are 6 good subband images that span frequencies from 886 to 1171 MHz, and another 6 that span frequencies from 1286 to 1681~MHz.

\begin{figure}[htb!]
\centering
\includegraphics[width=\textwidth]{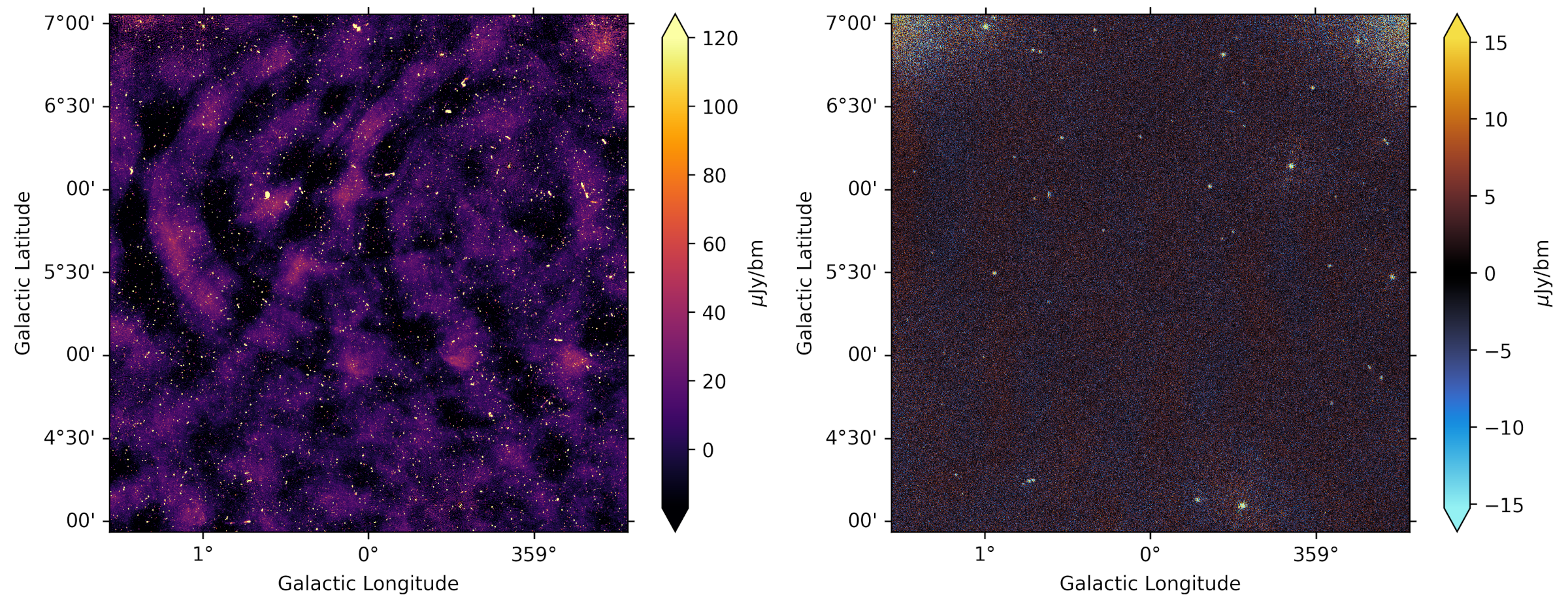}
\caption{A typical 3.125$^\circ$×3.125$^\circ$ mosaic field for one of the 20 pointings. These are Stokes I (left) and Stokes V (right) images for the G0.0+5.5 mosaic field. There are approximately continuum 25,000 radio sources. After eliminating image artifacts and other false positives (see \S\ref{sec:search}) we identify 4 known pulsars, 7 circularly polarized sources but no linear polarized sources in this mosaic field.}
\label{fig:mosaic}
\end{figure}

The 20 different mosaic fields given in the Galactic coordinate system ($l,b$), along with their properties are shown in Table \ref{tab:point} and plotted in Fig. \ref{fig:survey}.  In Fig.\ref{fig:mosaic} we show an example of the Stokes I (left) and Stokes V (right) mosaic images for the G0.0+5.5 pointing.

%\textcolor{red}{(Russ Taylor, PI of MIGHTEE) has suggested we might want to look at a mosaic weighting map. For our polarized sources, how many images went into making the mosaic?)}

%The Table will be more than just l,b. It will eventually contain the number of sources, the rms noise (average), Notes and WHAT ELSE? 
%The high brightness temperature of this part of the Milky Way adds substantial noise to the system reducing the sensitivity as a function of latitude. The average RMS noise in the mosaics centered  at b $\pm$2.5 was 21.2 $\mu$Jy beam at b$\pm$14.5, 13.2$\mu$Jy beam and at b$\pm$17.5, 14.2 $mu$Jy beam

\begin{deluxetable}{lrcchhcr}
\tablecaption{Observational Properties of Bulge Mosaics}\label{tab:point}
\tablehead{\colhead{$l$} & \colhead{$b$} & \colhead{\# of} & \colhead{rms} & \nocolhead{P T$_{sky}$} & \nocolhead{L T$_{sky}$} & \colhead{Known} & \colhead{Candi.}\\
\colhead{($\deg$)} & \colhead{($\deg$)}  & \colhead{Sources} & \colhead{($\mu$Jy/ba)} & \nocolhead{(K)} & \nocolhead{(K)} & \colhead{PSRs} & 
\colhead{(V, P)}}
\startdata
000.0	& +17.5	    & 21,532 & 14.6 & 52.5 & 4.1 & 1  & 1, 0\\
000.0	& +14.5	    & 23,088 & 13.9 & 60.9 & 4.4 & 1  & 0, 0\\
000.0	& +11.5	    & 17,287 & 16.9 & 79.4 & 4.7 & 0  & 2, 1\\
000.0	& +8.5	    & 17,966 & 17.7 & 85.4 & 4.9 & 2  & 3, 1\\
003.0	& +5.5	    & 15,819 & 16.0 &120.8 & 5.5 & 4  & 1, 2\\
357.0	& +5.5	    & 13,367 & 17.4 &121.8 & 5.7 & 3  & 4, 0\\
000.0	& +5.5	    & 25,086 & 12.5 &124.3 & 5.8 & 4  & 7, 0\\%+1 low-band
003.0	& +2.5	    & 17,769 & 14.7 &221.5 & 7.9 & 7  & 2, 0\\
357.0	& +2.5	    & 14,261 & 15.3 &251.5 & 8.6 & 9  & 0, 1\\
000.0	& +2.5	    & 18,815 & 14.3 &225.7 & 8.0 & 10 & 4, 2\\%+1 low-band
000.0	& $-$2.5	& 19,709 & 14.0 &206.1 & 7.6 & 18 & 6, 0\\
357.0	& $-$2.5    & 14,451 & 14.7 &214.8 & 8.2 & 7  & 3, 0\\	
003.0	& $-$2.5    & 12,715 & 16.8 &240.4 & 8.3 & 11 & 4, 0\\
000.0	& $-$5.5	& 29,992 & 10.7 &139.6 & 6.2 & 4  & 1, 0\\%-1 id=21770 not real
357.0	& $-$5.5	& 22,252 & 11.9 &122.1 & 5.7 & 5  & 3, 0\\%+1 low-band
003.0	& $-$5.5    & 21,019 & 11.9 &132.9 & 6.2 & 2  & 3, 0\\
000.0	& $-$8.5	& 23,350 & 13.4 & 94.6 & 5.0 & 0  & 6, 4\\%+1 low-band
000.0	& $-$11.5	& 20,847 & 13.5 & 69.3 & 4.8 & 1  & 4, 1\\
000.0	& $-$14.5	& 27,358 & 11.4 & 51.9 & 4.3 & 0  & 4, 0\\%+1 low-band
000.0	& $-$17.5	& 26,322 & 12.1 & 52.1 & 4.2 &1  & 9, 0\\
\enddata
%\tablecomments{Anything you need. Add a column for known pulsars?} 
\end{deluxetable}

\section{Source Catalog}\label{sec:catalog}

%{\it Emil: How did we form the catalog for each pointing.} Section 3.2 of Wang et al 2022 MNRAS, 516, 5972 has a nice description and Prichard et al reference on how they created the catalog.

For the source catalog and the following analysis we used a preliminary version of the mosaic images. For the entire survey, including full Stokes mosaic images and a gallery of extended sources, we refer the reader to \citep{cotton2024}. As our primary scientific goal is producing a list of pulsar candidates, the following discussions on source catalogs, their properties, and any data issues are focused on unresolved point sources.

We utilized PyBDSF \citep{2015ascl.soft02007M} to generate source catalogs for all images. PyBDSF subtracts a mean background map from the pixel intensities and normalizes each pixel value to the local noise, which is determined through interpolation from a coarsely sampled grid. Extensive testing showed a grid sampling of 25 pixels provides robust background subtraction and source flux measurements. The mean pixel intensity and standard deviation is calculated within a square box of 75 pixels on a side centered at each grid point. Pixels belonging to sources are excluded from the background calculation by iteratively rejecting outliers with large positive values. Islands of connected pixels that exceed 3 times the local noise are used to define sources, with elliptical Gaussian fitting initiated at the locations of pixel peaks within these islands. Sources with a signal-to-noise ratio (S/N) $<$ 5 and sources within 10 pixels of the image borders are discarded. We use the source catalogs of combined Gaussian components to separate mutli-component from single-component sources, a characteristic of unresolved point sources. Negative polarizations were cataloged by running PyBDSF on the negated Stokes Q, U, and V images.

Linear polarization images were created for each subband ($P = \sqrt{Q^2 + U^2}$) and for the full bandwidth by averaging $P$ over all subbands. Additionally, half bandwidth and quarter bandwidth images were constructed by averaging subbands in groups of 6 and 3, respectively, for all polarizations. Source spectra (S$_\nu \propto \nu^\alpha$) were independently determined in the half-band, quarter-band, and subband images through inverse-variance weighted fits to sources detected more than once. Note that we neglected to correct our linear polarization images for noise bias \citep{2012PASA...29..214G}. Originally we intended to focus our efforts exclusively on finding circularly polarized sources, but in the spirit of investigating the full capabilities of this dataset, we decided to add linearly polarized searches. The lessons learned from this pilot effort, including the impact of neglecting the noise bias, are discussed in more detail in \S\ref{sec:known} and \S\ref{sec:lin-candi}.

The presence of blanked regions in the mosaic subbands complicates determining source spectra in the half-band and quarter-band images. At lower frequencies, these blanked regions are randomly distributed across the field due to the need for RFI flagging of entire pointings. At higher frequencies, the field of view is smaller, and blanking tends to occur near the mosaic edges. To account for this, we calculated the average frequency across the subband-combined images, excluding the blanked regions. We then used these position-dependent frequencies for fitting source spectra. 

The full bandwidth mosaic's Stokes I catalog serves as the foundation for the complete source catalog in each field. Polarized sources were identified by locating the nearest source within 8 arcsec in the polarization catalogs. Spectral indices were assigned by matching the full bandwidth catalog to the half-band, quarter-band, and subband source catalogs.

To classify sources as compact or extended, we adopt the popular method \citep[e.g.,][]{2017A&A...598A..78I,2018MNRAS.474.5008D} analyzing the total-to-peak flux ratio ($R$) of single-component sources binned by S/N for each field catalog. We calculate the average and standard deviation of $R$ within each bin after iteratively removing $2\sigma$ outliers to exclude extended sources. We then fit a smoothly varying function of S/N to the $+2\sigma$ envelope using an empirical equation adapted from \citet{will2013}. We defined a compactness metric as the ratio of $R$ obtained from the fitted equation to the value measured for the source, serving as a statistical measure of source extent. Single-component sources with a compactness greater than 1 have a $\sim 97\%$ probability of being unresolved.

\subsection{Survey Properties}\label{sec:properties}

Table~\ref{tab:point} displays the number of sources in each field catalog. By associating sources within 8 arcsec across different fields, we identified 387,875 unique sources within the 173~deg$^2$ survey area. Among these, 375,286 are classified as single-component, with 298,785 being unresolved (compactness $> 1$). 188,753 sources lack in-band spectral index determinations. For circularly polarized sources without spectral indices (\S\ref{sec:candi}), we utilize PyBDSF in forced-fitting mode where the full-band source position is used to initiate Gaussian fitting in the individual subbands and subband combination images, enabling spectral indices to be obtained with flux measurements of weaker detections (S/N $> 3$).

% * minimum nearest neighbor distance in field catalogs is 5.4 arcsec
% * 387875 unique sources, using < 5 arcsec matching separation to avoid false associations
% * 349765 (of 403004) are S
% * 279841 (of 403004) are S and compactness > 1
% * 389795 (of 403004) are S or C
% * 309657 (of 403004) are S or C and compactness > 1
% * 13209 (of 403004) are M
% * 336717 (of 387875) are S
% * 269993 (of 387875) are S and compactness > 1
% * 375286 (of 387875) are S or C
% * 298785 (of 387875) are S or C and compactness > 1
% * 12805 (of 387875) are M
% * 194476 (of 403004) have no spix (6ave, 3ave, or noave)
%    * 193164 of these are S or C
% * 188753 (of 387875) have no spix
%    * 187474 of these are S or C

The number of International Celestial Reference Frame \citep[ICRF,][]{icrf3} sources within our survery area is insufficient for direct astrometric correction calculation. To tackle this issue, we employed a two-step correction process. First, we corrected each field to the RACS-mid catalog \citep{2023PASA...40...34D}, utilizing the median offset of unresolved sources isolated from their nearest neighbors by at least 16 arcsec and within 8 arcsec of an unresolved RACS-mid source. We achieved $235-369$ matches per field, with median offsets and standard deviations of $\Delta \alpha$ $\cos \delta = $  ($-0.33$ - $1.02$) $\pm$ ($0.50$ - $0.85$) arcsec, and $\Delta \delta = $ ($-0.53$ - $1.43$) $\pm$ ($0.43$ - $0.92$) arcsec. The astrometric offset magnitudes are less than 1 pixel for all fields. Second, we determined the correction to ICRF using the offset of unresolved RACS-mid sources within 8 arcsec of ICRF sources. To account for potential systematic effects, we limited our calculation to RACS-mid sources within 40 degrees of the GC. From 134 matches, we obtained $\Delta \alpha$ $\cos \delta = -0.21 \pm 0.35$ arcsec and $\Delta \delta = 0.10 \pm 0.31$ arcsec. These offsets are within 1$\sigma$ of those calculated for the entire catalog \citep{2023PASA...40...34D}.

\begin{figure}[htb]
    \centering
    \includegraphics[width=4in]{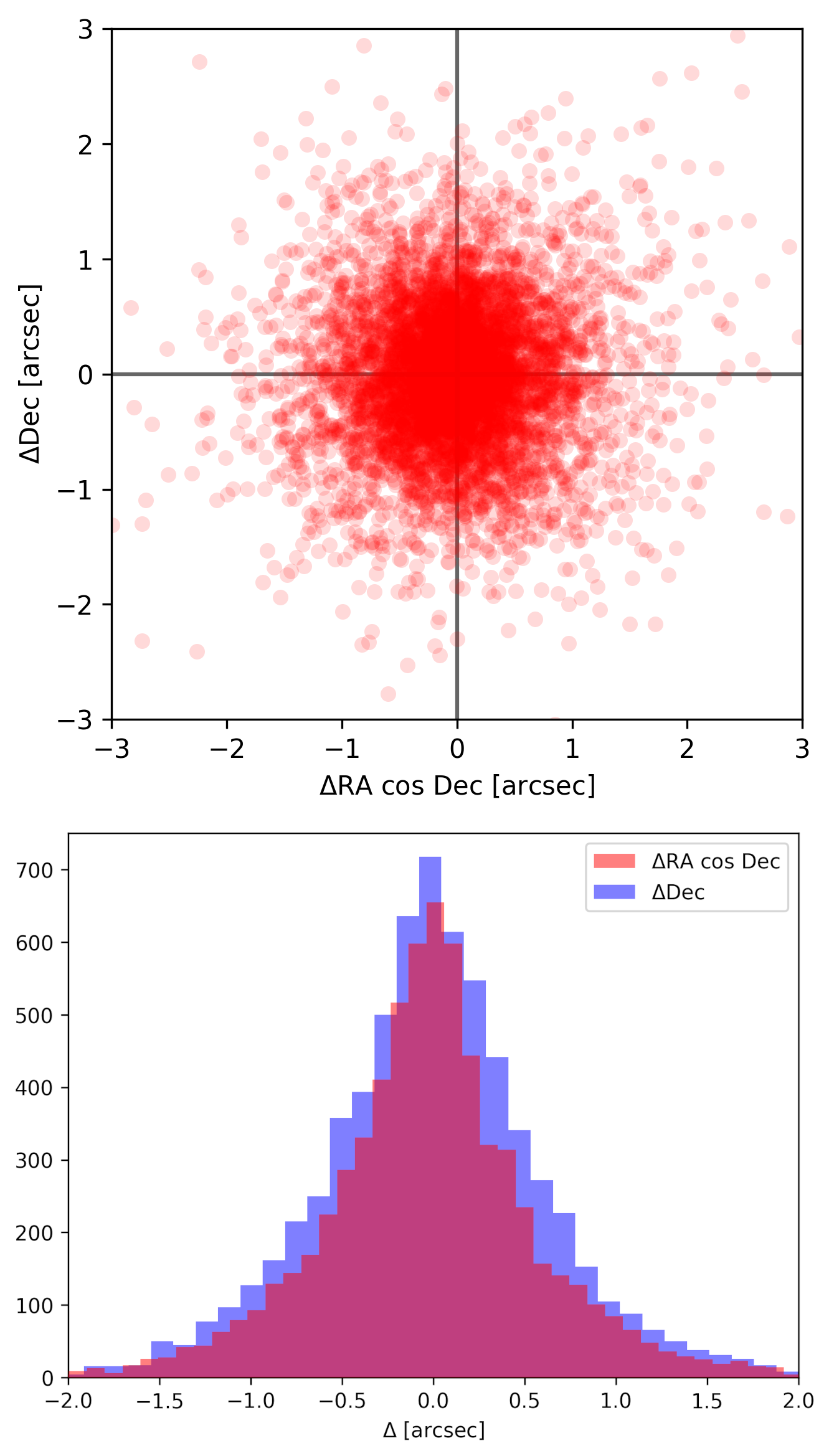}
    \caption{Angular offsets in right ascension and declination for 6766 unresolved sources detected in both this survey and the RACS-mid, after applying astrometric corrections. See \S\ref{sec:properties} for details.} \label{off}
\end{figure}

Figure~\ref{off} shows the offset distribution for all 6766 sources with RACS-mid matches, after correcting for the median offset. The standard deviation is 0.65 arcsec in both coordinates for the entire sample.

%The GP survey was taken around the same time as this bulge survey (2018 Jul 21 and 2020 Mar 14). There is a very detailed discussion about astrometric errors in \S{4.4} of \cite{2023arXiv231207275G}. They use ICRF and the VLA CORNISH survey for astrometric comparisons. It would be useful to understand from Bill whether the systematic timing and frequency effects shown in their Figure 5 are also in our data, and whether they have been corrected in the mosaics. The median offset in ($l,b$ is 0.16,0.30 arcsec with a standard deviation of 0.5 arcsec. Look at their Figure 6 for the scatter plot.

%Astrometry offsets in arcsec for early MIGHTEE (data taken in 2018) catalog using MeerKAT/VLA comparisons are $-0.25,-0.08$ for the COSMOS field and $-0.20,-0.43$ for the XMM-LSS field \citep{2022MNRAS.509.2150H}. In a more recent paper that seems to include the full MIGHTEE dataset from 2018 and data from 2020, they recompute the astrometric offsets for all 4 fields \citep{2024MNRAS.528.2511T} and get offsets of order 0.25 arcsec and 0.55 arcsec in RA and Dec. The COSMOS field is notable in having a much smaller error in declination offsets (0.14 arcsec). Their Figure 13 and Table 5 are well worth looking at.

%and later on \citep{2024MNRAS.527.3231W}

\subsection{Known Data Issues}\label{sec:issues}

These data were taken during an active development and commissioning stage of MeerKAT and thus there are a number of potential instrumental and calibration issues (e.g., labeling errors, calibrator position errors, delay tracking errors). These errors are detailed in \citet{2022ApJ...925..165H} but at the time of these observations the dominant error was the delay tracking error; basically the existing correlator model did not accurately transfer phases when switching between calibrator and target. In \S\ref{sec:properties} we looked for systematic errors in the astrometry and attempted to correct for the bulk of these uncertainties. Our resulting offsets and rms deviations in source positions are comparable to similar values estimated for other datasets taken around the same time and using similar methods \citep{2022ApJ...925..165H,2023arXiv231207275G,2024MNRAS.528.2511T}. 

\begin{figure}[htb]
    \centering
    \includegraphics[width=4in]{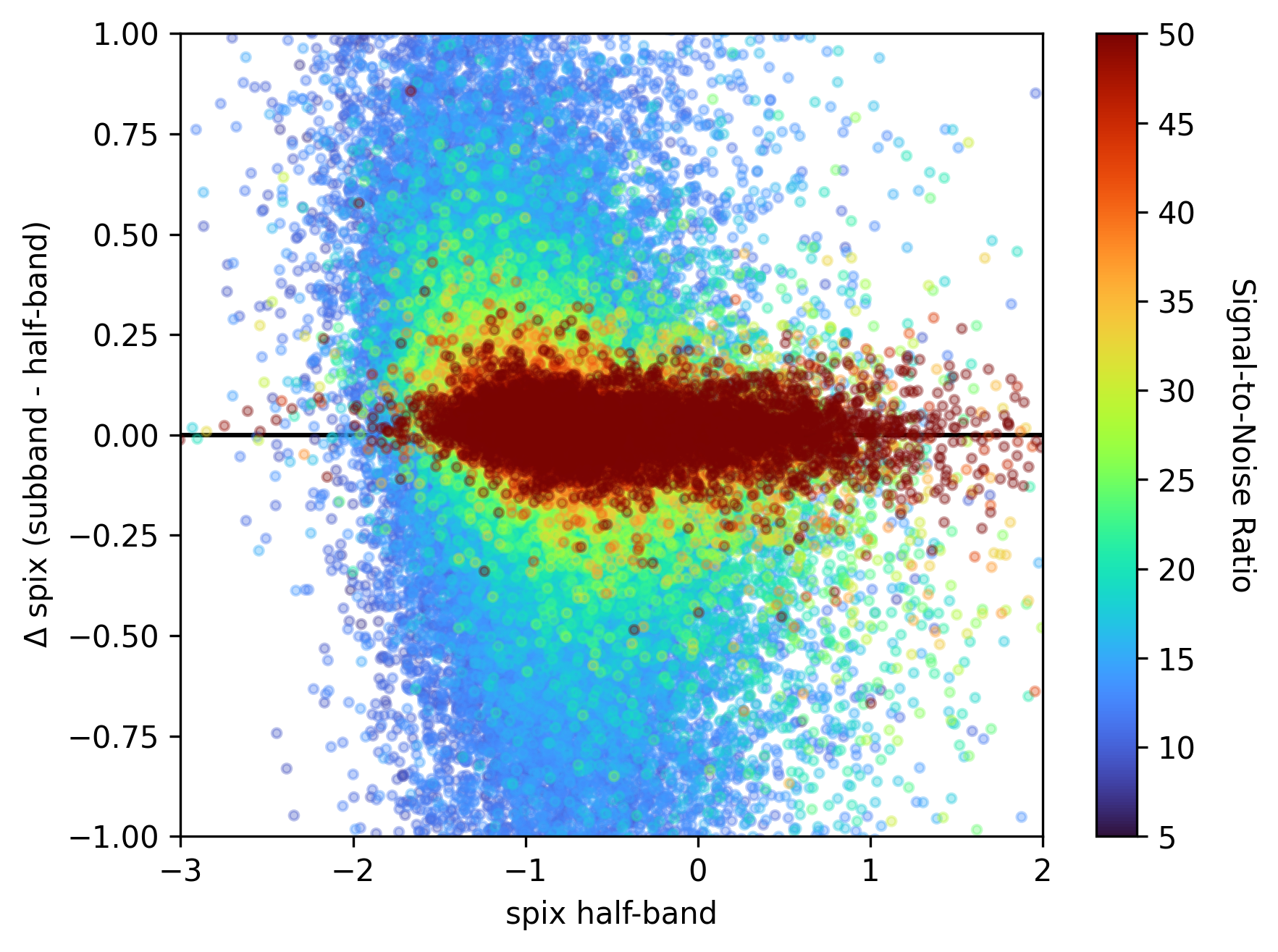}
    \caption{The difference in source spectral index derived using the two half-band averaged images and the individual 12 subband images for 85,273 compact, single component sources. Sources are colored according to their S/N in the full-band image.}  \label{fig:spix_compare}
\end{figure}

In order to improve S/N of the in-band spectral indices we averaged different subbands together in six and three adjacent frequency bands (see \S\ref{sec:catalog}). To test the reliability of this averaging method we compared these derived spectral indices with those fitted using all 12 subbands. This is shown in Fig. \ref{fig:spix_compare} where we compare the difference between the un-averaged subbands and the half-band averages (6 adjacent frequencies) for 85,273 compact, single component sources. The difference between these two spectral index values is within the best-fit errors for 73\% of these compact, single-component sources, with the larger scatter occurring for low S/N sources. In our search for known and candidate pulsars below we primarily used the $\alpha$ values formed from the six sub-band averages. However, when we saw significant deviations between the other measures of $\alpha$, we visually inspected the un-averaged spectra. The differences were usually due to RFI or low S/N resulting in a poor fit from PyBDSF. In those few cases we simply manually re-fit the data but there was an interesting subset of steep spectrum sources that were too faint at higher frequencies that required a different search strategy (see \S\ref{sec:lowhangingcandi}). \citet{2023arXiv231207275G} and \citet{2024MNRAS.529.2443C} have detailed other issues related to the derivation of source spectral indices related to the lack of short spacings in these Galactic fields. This leads to a frequency dependent ``zero offset" error that can steepen $\alpha$ from its true value. As this project is concerned with unresolved point sources, it should not be a serious problem unless the source is embedded in diffuse emission. Such cases can be identified by inspecting the images directly.

Interferometers like MeerKAT with orthogonal linearly polarized feeds have other instrumental effects. The most notable for polarimetric imaging is due to antenna pointing errors, which result in a frequency-dependent instrumental polarization signal also called ``beam squint" \citep{2022AJ....163...87S}. \cite{2024MNRAS.528.2511T} have shown that the polarization leakage for MeerKAT is negligible ($<0.2$\%) over most of the L-band frequency range, but rises to several percent for frequencies above 1.4 GHz for sources whose position is offset more than 0.45$^\circ$ from the pointing center of the primary beam. For this particular project there is an added effect that reduces beam squint. We are looking for pulsars whose typically steep spectra ($\alpha\simeq{-1.7}$) weight down the instrumental polarization by a factor of three at higher frequencies compared to the lower frequencies. Some bright, {\it flat spectrum} sources with values of polarization of order 1-2\% could still be instrumental artifacts, but in any case their spectra would not make them promising pulsar candidates.

\begin{figure}[htb]
    \centering
    \includegraphics[width=3in]{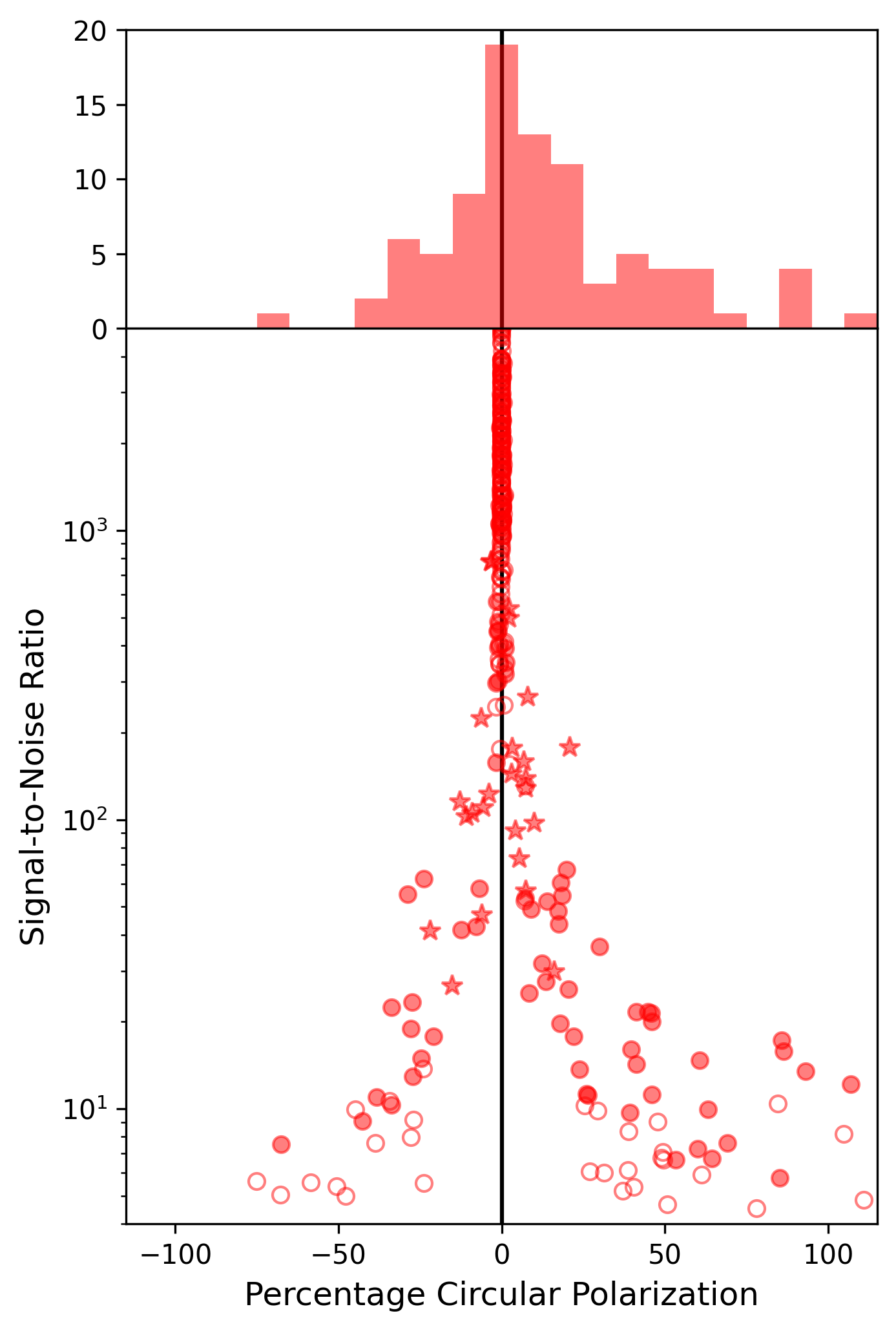}
    \caption{Polarized flux fraction distribution for 353 circularly polarized sources. Star symbols indicate known pulsars. Open circles are identified as image artifacts in \S\ref{sec:candi} and not included in the upper histogram plot.} \label{fig:vfrac}
\end{figure}

In Fig.\,\ref{fig:vfrac} we plot all 353 sources with detectable circular polarization as a function of Stokes I S/N. As we discussed above, we see a large number of detections at the 1\% level that are likely instrumental artifacts arising from polarization leakage. For this reason bright Stokes I sources with $\vert{\rm V/I}\vert<1$\% were eliminated 
from the sample. This value is in line with other estimates of polarization leakage from recent ASKAP and LOFAR polarimetric imaging surveys \citep{2022MNRAS.516.5972W, 2023A&A...670A.124C}. We will discuss the additional image artifacts (open circles) and the histogram in \S\ref{sec:candi}.

There is a positive skew in the distribution of V/I sources (Fig. \ref{fig:vfrac})
which is most pronounced at low Stokes I S/N ($<$30) and high degrees of polarization  ($\vert${\rm{V/I}}$\vert$$>$30\%). While certain coherent mechanisms favor amplification of the x-mode, which increase the {\it degree} of polarization \citep[see ][]{1992A&A...264L..31G}, we know of no physical mechanism that would favor the {\it handedness} of the polarization. The origin of this excess is also unlikely due to the data calibration and deconvolution process, since the number of positive and negative Stokes V sources are very similar in each of the mosaic fields.

We do not appear, however, to be overestimating the number of positive V/I sources. If this sample is dominated by false positives, we might expect that the fraction of  multi-wavelength counterparts for positive V/I sources to drop compared to negative V/I sources at higher fractional polarization or lower signal-to noise. As we note in \S\ref{sec:finalcandi} this appears not to be the case. Thus we appear to be under counting the number of real negative V/I sources. We 
have been unable to find a systematic error in our process that would under count negative compact polarized radio sources.  We followed a standard process to identify and then catalog the Stokes parameters for all significant sources in the mosaic images (\S\ref{sec:catalog}) that is not biased against matching negative V Stokes with their Stokes I counterparts. Since we used a preliminary release of the MeerKAT images, it will be instructive to see if this asymmetry persists in the final public data release \citep{cotton2024}.

\section{Polarization Candidate Search}\label{sec:search}

\subsection{Known Pulsars}\label{sec:known}

The efficacy of this survey for finding pulsar candidates can be tested using the sample of known pulsars. There are 81 known pulsars that lie within the survey area (see Fig. \ref{fig:survey} and Table \ref{tab:point}). This is an under count of the true number of pulsars.
There are two globular clusters (NGC\,6522 and Terzan 5) that collectively have at least 45 pulsars. However, unless the pulsar is well separated from the cluster core, our images lack the angular resolution to resolve them individually,  \citep{2020ApJ...904..147U,2020ApJ...905L...8Z,2023A&A...680A..47A}. We identified three discrete, polarized point sources in these clusters for inclusion in Table \ref{tab:known}. We suspect that the polarized radio source identified as PSR\,J1748$-$2446ab may be an amalgamation of several overlapping pulsars.

For each mosaic pointing we first cross-matched our source catalog with the ATNF pulsar catalog\footnote{ http://www.atnf.csiro.au/research/pulsar/psrcat} \citep{2005AJ....129.1993M}.  When possible we substituted ATNF catalog positions with newer values from the literature. For example, the identification of the RRAT PSR\,J1739$-$2521 used a position from \cite{2017ApJ...840....5C}. For Stokes I, this resulted in nearly a 100\% detection rate. The quality of the pulsar position errors in this sample varies considerably; with sub-arcsec accuracy for well-timed MSPs, to arcminute positions for RRATs. As a result, and given our superb continuum sensitivity, it is likely that identifying pulsars relying on {\it only} Stokes I would result in large number of detections of unrelated background extragalactic radio sources. To reduce these false positives, we required a positive pulsar match to also have either linear or circular polarization. 

Table \ref{tab:known} lists the properties of the 39 known {\it polarized} pulsars detected in this survey. Circular polarization is listed under the column V/I, while linear polarization is listed under P/I; S/N of each value is listed in the adjacent column. Our failure to correct our linear polarization images for a well-known noise bias (see \S\ref{sec:catalog}), has a minor effect on the P/I values. We estimated its effect by propagating the errors by hand, including estimates for the noise bias, for a small number of pulsars in this table. We found that the estimates for the lower S/N values in the table were overestimated by 20-40\% in some cases. Nevertheless, we verified from visual inspection of the images that all of the detections of linear (and circular) polarization in Table \ref{tab:known} are real detections and not image artifacts.

In a number of cases the uncertainties on our measured positions (RA/DEC), flux density (I), spectral indices ($\alpha$) and polarization values are significant improvements over what is available in the literature. For the majority of the cases our positions are consistent with the known pulsar positions within joint measurement uncertainties, with a few exceptions. Specifically, the radio source identified with the MSP PSR\,J1653$-$2054 is 4-arcsec from its timing position, while for PSRs J1720$-$2446 and J1751$-$3323 the timing vs interferometic positions are offset by 1.5$\sigma$ and 2$\sigma$, respectively. It is unlikely that these steep spectrum, polarized point sources are false pulsar identifications as such sources are rare. We show in the next section that the source density of circularly polarized sources is only 0.5 deg$^{-2}$, and \citet{2018MNRAS.474.5008D} showed that $<1$\% of radio sources have $\alpha<-1.5$. Proper motion might explain the offsets but there are no published values in the literature. It seems likely that unmodeled timing residuals or optimistic position error estimates could account for the two normal pulsars, but the 4-arcsec offset for the MSP is difficult to explain.

The peak flux densities (I) of the pulsars in Table \ref{tab:known} range from about 0.4 to 40 mJy, and are broadly consistent with the values in the ATNF catalog. The spectral indices $\alpha$ range from +0.07 to $-3.5$, with a median value of $-1.8\pm{0.8}$, in reasonable agreement with the weighted mean spectral index of 441 pulsars of $-1.60\pm{0.03}$ \citep{2018MNRAS.473.4436J} and from the more recent value of $-1.78\pm{0.6}$ from 168 pulsars \citep{2023ApJ...956...28A}.  Nearly all of the pulsars have significant linear polarization, while less than 2/3 have significant circular polarization. The maximum linear polarization in Table \ref{tab:known} is 45\%, while the maximum circular polarization is about half this amount. This range in polarization and the absolute values of the polarization percentages are consistent with the pulse-weighted averages from a large sample of pulsars \citep{2018MNRAS.474.4629J,1979ApJ...228..755R,2023ApJ...956...28A}. From the \citet{2018MNRAS.474.4629J} sample of 600 pulsars, the mean linear polarization is 27\% with a range from 1-100\%, while the mean circular polarization is 8.4\%, with a range from 1-40\%, with rare cases in excess of 60\%. Moreover, there are 18 pulsars in Table \ref{tab:known} which are also found in common with \citet{2018MNRAS.474.4629J}. We detect linear and/or circular polarization of the right magnitude from all but their two faintest pulsars. The {\it direction} of polarization is reversed in these two samples, suggesting different conventions between MeerKAT polarization and the known pulsar sample.

Summarizing, we find polarized point sources coincident with approximately 50\% of the known pulsars, and their properties (flux density, spectral index and polarization) are consistent with these radio sources as known pulsars. 

%Results. Some topics for discussion of known pulsars (1) Basic summary of the results. How many, etc.What fraction of detections are polarized.  (b) Using polarized rather than I as criteria because of the source density. What is the radius before random associations? (c) Offsets. Why are there ones with larger offsets than timing errors. (d) Missing pulsars that might be in our polarization sample due to timing errors. Two cases so far. (e) Ter 5. More detail.. Did we detect J1748-2446aa or CXO J174804.7-244642? (f) How do properties of known pulsars compare to our candidates (for the later discussion of candidates)?

%\begin{splitdeluxtable}{lrrllBrrrrllr}
\begin{rotatetable}
\begin{deluxetable}{lrrllrrrrllh}
\tabletypesize{\scriptsize}
\tablewidth{0pt} 
\tablecaption{The Polarization Properties of Known Pulsars\label{tab:known}}
\tablehead{\colhead{PSR Name} & \colhead{P} & \colhead{DM} & \colhead{RA} & \colhead{DEC} & \colhead{I} & \colhead{$\alpha$} & \colhead{{V/I}} &  \colhead{S/N} & \colhead{P/I} &  \colhead{S/N} & \nocolhead{Notes}\\
\colhead{} & \colhead{(ms)} & \colhead{(pc cm$^{-3}$)} & \colhead{(h~~m~~s)} & \colhead{($^\circ$\ ~~$'$\ ~~$"$)} & \colhead{(mJy)} & \colhead{} & \colhead{(\%)} & \colhead{} & \colhead{(\%)} & \colhead{} & \nocolhead{}
}
\startdata
J1653$-$2054 & 4.1 & 56.5  & 16:53:30.79 & $-$20:54:58.1 & 2.19$\pm$0.01 & $-$1.33$\pm$0.02 &  \omit & \omit & 1.7 & 6.2 & g0.0+14.5 11473 4.2\\
J1719$-$2330 & 454.0 & 101.0  & 17:19:36.44 & $-$23:30:07.3 & 2.89$\pm$0.09 & $-$1.16$\pm${0.44} &  \omit & \omit & 3.8 & 6.9 & g0.0+8.5 274 8.2\\
% Duplicate. J1720$-$2446 & 874.3 & 104.3  & 17:20:22.58 & $-$24:46:09.1 & 0.74$\pm$0.05 & \omit &  \omit & \omit & 8.1 & 6.9 & g0.0+8.5 6045 18.0\\
J1720$-$2446 & 874.3 & 104.3  & 17:20:22.57 & $-$24:46:09.0 & 0.45$\pm$0.02 & $-$2.52$\pm$0.19 &  \omit & \omit & 14.1 & 7.1 & g0.0+5.5 8791 18.1\\
B1717$-$29 & 620.4 & 42.6  & 17:20:34.11 & $-$29:33:16.1 & 2.68$\pm$0.01 & $-$1.16$\pm$0.03 &  $-$6.3 & 19.4 & 9.5 & 22.8 & g357+5.5 12275 0.2\\
J1721$-$2457 & 3.5 & 48.2  & 17:21:05.50 & $-$24:57:06.7 & 1.37$\pm$0.01 & $-$1.17$\pm$0.05 &  $-$12.8 & 23.5 & 5.0 & 10.8 & g0.0+5.5 9278 1.4\\
%Duplicate. J1727$-$2739 & 1293.1 & 146.0  & 17:27:31.02 & $-$27:38:51.8 & 1.84$\pm$0.01 & $-$1.47$\pm$0.04 &  7.4 & 17.0 & 35.3 & 50.4 & g0.0+5.5 20696 2.0\\
J1727$-$2739 & 1293.1 & 146.0  & 17:27:31.03 & $-$27:38:53.0 & 1.76$\pm$0.01 & $-$1.50$\pm$0.04 &  7.3 & 15.4 & 37.1 & 48.8 & g0.0+2.5 15772 1.7\\
J1730$-$2304 & 8.1 & 9.6  & 17:30:21.69 & $-$23:04:31.3 & 4.36$\pm$0.02 & $-$0.31$\pm$0.03 &  20.9 & 86.4 & 21.6 & 120.6 & g3.0+5.5 3989 1.2\\
B1730$-$22 & 871.7 & 41.1  & 17:33:26.41 & $-$22:28:45.3 & 12.97$\pm$0.09 & $-$3.10$\pm${0.18} &  3.1 & 12.5 & 12.6 & 61.9 & g3.0+5.5 110 8.3\\
% This next one was from the linear candidate list 6982 but was missed because it has +/-90 arcsec dec timing error
J1734$-$2415 & 612.5 & 126.3 & 17:34:41.37 &  $-$24:16:15.0 & 0.60$\pm$0.01 & $-$1.45$\pm$0.10 & \omit & \omit & 46.8 & 30.2 & g3.0+5.5 6982\\
J1736$-$2457 & 2642.2 & 169.4  & 17:36:45.37 & $-$24:58:02.4 & 1.06$\pm$0.01 & $-$2.14$\pm$0.05 &  \omit & \omit & 24.3 & 19.4 & g3.0+2.5 11438 12.4\\
B1736$-$29 & 322.9 & 138.6  & 17:39:34.28 & $-$29:03:02.7 & 4.86$\pm$0.04 & $-$1.55$\pm$0.07 &  \omit & \omit & 4.2 & 10.1 & g0.0+2.5 14275 1.1\\
% Next line is from the linear candidate list 11384. RRAT with a new position from Cui, Boyles, McLaughlin et al ApJ, 840, 2017. 
J1739$-$2521 & 1818.5 & 186.4  & 17:39:32.79 & $-$25:21:09.3 & 0.35$\pm$0.01 & $-$2.54$\pm$0.16 & \omit & \omit & 15.9 & 6.2 & g3.0+2.5 11384 \\ 
J1741$-$2733 & 893.0 & 147.4  & 17:41:01.33 & $-$27:33:56.2 & 1.89$\pm$0.02 & $-$2.35$\pm$0.05 &  $-$5.7 & 10.1 & 17.0 & 23.7 & g0.0+2.5 5907 5.3\\
B1740$-$31 & 2414.7 & 193.1  & 17:43:36.72 & $-$31:50:22.5 & 2.09$\pm$0.02 & $-$2.54$\pm$0.05 &  4.2 & 8.2 & 13.7 & 32.4 & g357$-$2.5 7760 0.3\\
J1743$-$3153 & 193.1 & 505.7  & 17:43:15.59 & $-$31:53:05.8 & 0.73$\pm$0.02 & +0.07$\pm$0.15 &  $-$22.0 & 14.0 & 23.1 & 17.7 & g357$-$2.5 8207 1.3\\
J1744$-$3130 & 1066.1 & 192.9  & 17:44:05.71 & $-$31:30:06.5 & 0.74$\pm$0.02 & $-$1.37$\pm${0.05} &  \omit & \omit & 13.3 & 10.1 & g357$-$2.5 5753 2.8\\
B1742$-$30 & 367.4 & 88.4  & 17:45:56.32 & $-$30:40:22.9 & 18.85$\pm$0.04 & $-$1.87$\pm$0.10 &  2.2 & 27.2 & 43.4 & 530.1 & g0.0$-$2.5 18973 0.1\\
%Duplicate. B1742$-$30 & 367.4 & 88.4  & 17:45:56.33 & $-$30:40:22.9 & 18.25$\pm$0.03 & \omit &  2.2 & 27.2 & 44.2 & 542.9 & g357$-$2.5 58 0.4\\
B1744$-$24A & 11.6 & 242.1  & 17:48:02.26 & $-$24:46:37.7 & 3.45$\pm$0.04 & $-$2.48$\pm$0.11 &  \omit & \omit & 4.0 & 6.0 & g3.0+2.5 636 1.2\\
J1748$-$2446ab & 5.1 & 242.2  & 17:48:04.74 & $-$24:46:41.8 & 0.9$\pm$0.04 & \omit &  \omit & \omit & 19.0 & 7.2 & g3.0+2.5 612 3.3\\
J1748$-$3009 & 9.7 & 420.2  & 17:48:23.75 & $-$30:09:11.6 & 2.19$\pm$0.17 & $-$2.25$\pm$0.38 &  $-$7.1 & 6.8 & 16.6 & 19.8 & g0.0$-$2.5 14844 0.4\\
B1746$-$30 & 609.9 & 509.4  & 17:49:13.51 & $-$30:02:36.4 & 4.84$\pm$0.05 & $-$1.99$\pm$0.06 &  $-$10.9 & 26.8 & 11.3 & 29.8 & g0.0$-$2.5 13687 1.0\\
B1747$-$31 & 910.4 & 206.3  & 17:50:47.31 & $-$31:57:44.5 & 1.64$\pm$0.01 & $-$1.58$\pm$0.03 &  3.3 & 7.6 & 27.6 & 46.1 & g357$-$2.5 3522 0.4\\
J1751$-$2857 & 3.9 & 42.8  & 17:51:32.65 & $-$28:57:47.1 & 0.85$\pm$0.07 & $-$2.42$\pm$0.75 &  \omit & \omit & 37.0 & 14.6 & g0.0$-$2.5 6357 0.5\\
J1751$-$3323 & 548.2 & 296.7  & 17:51:32.76 & $-$33:23:34.5 & 1.59$\pm$0.01 & $-$0.59$\pm$0.03 &  \omit & \omit & 15.4 & 40.2 & g357$-$2.5 10511 5.1\\
B1749$-$28 & 562.6 & 50.4  & 17:52:58.67 & $-$28:06:38.3 & 39.22$\pm$0.05 & $-$3.01$\pm$0.01 &  $-$3.2 & 73.9 & 11.2 & 267.7 & g0.0$-$2.5 141 0.7\\
%B1749$-$28 & 562.6 & 50.4  & 17:52:58.69 & $-$28:06:38.2 & 38.49$\pm$0.05 & $-$3.01$\pm$0.01 &  $-$3.3 & 74.0 & 11.4 & 278.4 & g3.0$-$2.5 11998 1.5\\
J1754$-$3443 & 361.7 & 188.7  & 17:54:37.34 & $-$34:43:55.0 & 1.01$\pm$0.03 & $-$1.63$\pm$0.59 &  16.2 & 6.2 & 10.3 & 5.6 & g357$-$5.5 22060 1.2\\
J1759$-$2922 & 574.4 & 79.4  & 17:59:48.26 & $-$29:22:08.1 & 0.83$\pm$0.01 & $-$2.61$\pm$0.07 &  5.5 & 5.3 & 19.9 & 21.7 & g0.0$-$2.5 2583 0.7\\
J1759$-$3107 & 1079.0 & 128.3  & 17:59:22.05 & $-$31:07:22.0 & 1.31$\pm$0.01 & $-$1.49$\pm$0.04 &  $-$3.9 & 6.7 & 16.8 & 27.1 & g0.0$-$2.5 12638 0.3\\
B1758$-$29 & 1081.9 & 125.6  & 18:01:46.82 & $-$29:20:39.6 & 2.95$\pm$0.01 & $-$1.79$\pm$0.02 &  8.0 & 31.6 & 43.5 & 170.4 & g0.0$-$2.5 887 0.7\\
J1801$-$3210 & 7.5 & 177.7  & 18:01:25.90 & $-$32:10:53.8 & 0.67$\pm$0.01 & $-$2.15$\pm$0.10 &  \omit & \omit & 13.0 & 13.7 & g0.0$-$5.5 25495 1.2\\
B1800$-$27 & 334.4 & 165.5  & 18:03:31.70 & $-$27:12:05.2 & 1.41$\pm$0.02 & $-$1.27$\pm$0.09 &  7.5 & 8.4 & 33.4 & 46.7 & g3.0$-$2.5 1686 1.6\\
% added from circ pol list. Pretty sure it is "A" in NGC 6522. See timing solution from Zhang, Manchester et al 2020.
J1803$-$3002A & 7.1 & 192.6  & 18:03:35.11& $-$30:02:00.8 & 0.66$\pm$0.01 & $-$2.15$\pm$0.09 & $-$6.9 & 5.8 & \omit & \omit & g0.0$-$2.5 3845 NGC\,6522\\
J1803$-$3329 & 633.4 & 170.9  & 18:03:44.45 & $-$33:29:10.9 & 0.92$\pm$0.01 & $-$1.26$\pm$0.05 &  $-$9.1 & 13.9 & 8.3 & 13.4 & g357$-$5.5 5319 0.7\\
J1804$-$2717 & 9.3 & 24.7  & 18:04:21.15 & $-$27:17:32.2 & 0.69$\pm$0.03 & $-$3.50$\pm$0.18 &  $-$15.2 & 8.4 & 39.7 & 28.0 & g3.0$-$2.5 1616 1.1\\
J1804$-$2858 & 1.5 & 232.5  & 18:04:01.53 & $-$28:58:47.2 & 1.09$\pm$0.01 & $-$2.45$\pm$0.05 &  9.9 & 15.5 & 5.4 & 8.6 & g3.0$-$2.5 8878 0.7\\
B1804$-$27 & 827.8 & 313.0  & 18:07:08.50 & $-$27:15:03.7 & 1.54$\pm$0.03 & $-$3.19$\pm$0.22 &  $-$6.1 & 5.0 & \omit & \omit & g3.0$-$2.5 355 1.5\\
J1808$-$3249 & 364.9 & 147.3  & 18:08:04.51 & $-$32:49:32.9 & 1.65$\pm$0.01 & $-$1.10$\pm$0.03 &  6.8 & 16.6 & 8.8 & 21.5 & g0.0$-$5.5 24424 1.8\\
J1812$-$2748 & 237.0 & 104.0  & 18:12:40.59 & $-$27:48:04.5 & 0.41$\pm$0.04 & $-$0.42$\pm$1.40 &  \omit & \omit & 25.8 & 5.9 & g3.0$-$5.5 285 1.9\\
J1812$-$3039 & 587.5 & 141.4  & 18:12:44.90 & $-$30:39:22.2 & 0.46$\pm$0.01 & $-$2.19$\pm$0.12 &  \omit & \omit & 28.1 & 18.7 & g0.0$-$5.5 1234 1.2\\
\enddata
\tablecomments{The positions (RA/DEC), flux densities (I), spectral indices ($\alpha$) and polarization fraction {V/I} and {P/I} come from this work, while the period (P) and dispersion measure (DM) are taken from the ATNF Pulsar Catalog \citep{2005AJ....129.1993M}.} 
\end{deluxetable}
\end{rotatetable}
%\end{splitdeluxtable}

\subsection{Circularly Polarized Sources}\label{sec:candi}

For each mosaic field we carried out a search for circular polarized sources. We started by identifying single Gaussian component circularly polarized sources within the catalog made from each mosaic with $\vert{\rm V/I}\vert\geq $1\%. The source density of circularly polarized sources is only 0.5 deg$^{-2}$, so this approach was sufficient to identify {\it all} the radio sources with non-zero Stokes V, and sort them in order of decreasing fractional polarization. 

To complete the search, these remaining sources were visually inspected directly from the Stokes I and V images to look for suspect signals (see \S\ref{sec:issues}). The most common V Stokes artifacts were produced from sidelobe contamination from bright radio sources, while a smaller number were flagged as suspicious as they were found at the far edge of the antenna's field-of-view, or the Stokes V signal was offset from the Stokes I peak by a significant fraction of the synthesized beam. Of the 100 initial sources
from the catalog search, 38 were flagged as suspicious or image artifacts, another was a duplicate identification from two adjacent mosaics, and another was a known pulsar that was found due to an improved timing position \citep{2020ApJ...905L...8Z}. The V/I values for the artifacts (open circles) and for both the known pulsars (filled stars) and our sources (filled circles) are shown in Fig.\,\ref{fig:vfrac}. While the V/I values for the known pulsars in Table \ref{tab:known} are evenly distributed, the 
sources in Table \ref{tab:circ} are skewed toward more positive V/I values. As we discuss in \S\ref{sec:issues}, the positive V/I excess occurs at lower values of Stokes I S/N. The known pulsars are on average brighter than our remaining circular polarized sources, and thus do not show this skew.

%Other polarization samples exhibit a slight skew in positive V/I \citep[e.g.,][]{2018MNRAS.474.4629J}, but our sample has a nearly 2:1 ratio. 
 
\begin{figure}
%\epsscale{0.8}
% Note that the x range is 0.05 t0 45 and the y tange is 0.85% to 115%. The Y/X ratio is about 1.2. circles size is -1.0*spix_6ave-0.25
%forPaper_PSRwCANDI.dat and psr600polar.fits
% Always start with the fits file first so grey points donot overwrite the blue/red nes.
% 
\centering
\plottwo{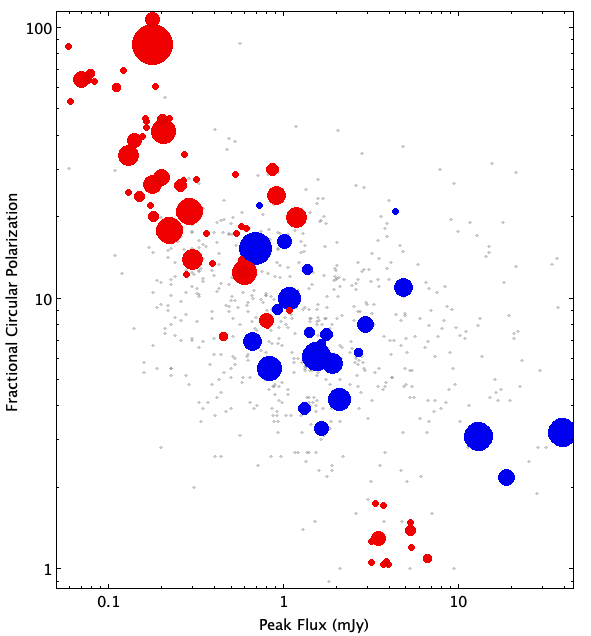}{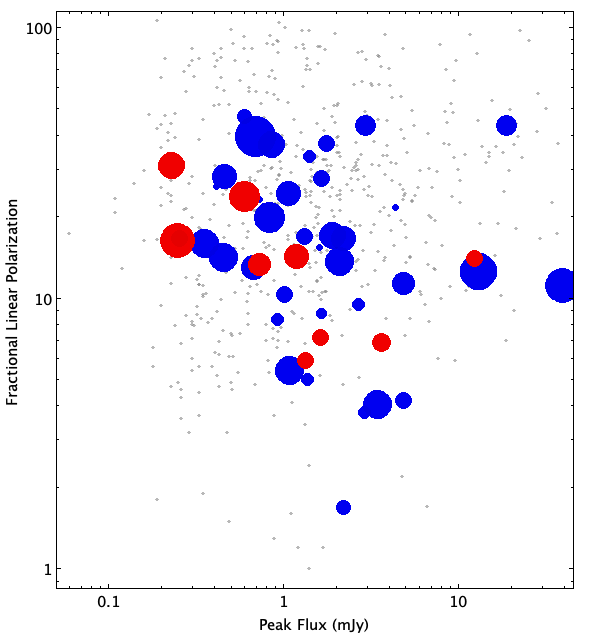}
\caption{Polarized sources (red) and known pulsars (blue). The left panel plots the absolute value of the circular polarization fraction versus Stokes I flux density, while the right plots the linear polarization versus Stokes I flux density. The size of the circle for each point is proportional to the spectral index, with larger circles indicating steeper negative spectral indices. Grey points are polarization and flux density values taken from the compilation of \citet{2018MNRAS.474.4629J}.}\label{fig:polar}
\end{figure}

 The ``V'' entries in column 6 of Table \ref{tab:point} list the number of circularly polarized sources found in each of the mosaic pointings.  Our final sample of 60 circularly polarized sources is given in Table \ref{tab:circ} and is plotted in Fig. \ref{fig:polar} (left). While the number of known pulsars in Fig. \ref{fig:survey} increases with decreasing Galactic latitude as expected, the distribution of circular polarized sources in Table \ref{tab:point} is more uniform, with clusters of sources 
 at high latitude (e.g., G000.0$-$17.5). This is suggestive that not all of these circular polarized sources are pulsars, but rather some other source population(s). This hypothesis is further supported by the source distribution in Fig.\,\ref{fig:polar}. The mean spectral index of our 60 sources is $-0.6$, versus $-1.8$ for the known pulsars in Table \ref{tab:known}. Likewise, the fractional polarization for the known pulsars in these fields ranges from 2-20\%, while our sources are distributed over a much larger range. 

The absence of sources in the bottom left hand corner of Fig.\,\ref{fig:polar} is due to a noise bias, the result of cutting off low S/N detections \citep[see also][]{2021MNRAS.502.5438P,2022MNRAS.516.5972W,2022A&A...661A..87S,2023A&A...670A.124C}.  The fixed MeerKAT integrations introduce a strong bias in our polarization measurements. For example, our sources must have a peak Stokes I flux in excess of $\sim$1 mJy to have a measured V/I$<$5\%. As this is rather bright, we expect that most such pulsars will have been found by previous time-domain surveys, while our sources on average will be fainter than that. The {\it real} distribution of pulsar circular polarization with peak flux is closer approximated by the large sample of \citet{2018MNRAS.474.4629J} plotted in  Fig.\,\ref{fig:polar} as light grey points. In that sample, there is no noise bias because the integration times for each of the pulsars are determined in order to achieve a S/N sufficient to measure their polarization properties. 

The lack of bright, highly polarized sources, however, is real (i.e, upper right corner). %Several of the highly polarized sources in the upper left corner are too faint to have a measured spectral index, and thus appear as a small diameter circle on this plot. 
As we noted in \S\ref{sec:issues}, the cluster of bright ($>$3 mJy) weakly polarized (1-2\%) sources in the bottom right corner could still have some false positives among them. While there are some steep spectrum sources in this low polarization sample that may have been missed by previous pulsation searches, we suspect that most are extragalactic sources with weak circular polarization or beam squint artifacts. There is support for this hypothesis since the median spectral index of this small sample more closely resembles that of extragalactic sources than pulsars \citep{2018MNRAS.474.5008D}. In \S\ref{sec:count} we will use multi-wavelength counterparts in an effort to distinguish promising pulsar candidates from other source populations.

\startlongtable
\begin{deluxetable}{llllrrrll}
\tablecaption{Circularly Polarized Sources\label{tab:circ}}
\tablehead{\colhead{RA} & \colhead{DEC} & \colhead{$l$} & \colhead{$b$} &  \colhead{I} & \colhead{$\alpha$} & \colhead{V/I} &  \colhead{S/N} & \colhead{Notes}\\
\colhead{(h~~m~~s)} & \colhead{($^\circ$\ ~~$'$\ ~~$"$)} & \colhead{(deg)} & \colhead{(deg)} & \colhead{(mJy)} & \colhead{} & \colhead{(\%)}  & \colhead{} & \colhead{}
}
\startdata
16:45:59.27 &  $-$18:22:20.1 &  0.96 &17.21 & 0.39$\pm$0.01 & 0.53$\pm$0.3 & 13.5 & 7.0 & O,IR; YSO\\%g0.0+17.5 3187\\
16:56:48.02 &  $-$23:09:01.8 &  358.6 &12.31 & 0.81$\pm$0.02 & 0.21$\pm$0.16 & $-$7.9 & 5.2 & O,IR; A05 star \\%g0.0+11.5 16811\\
17:00:27.74 &  $-$22:07:42.7 &  359.96 &12.24 & 0.08$\pm$0.01 & -1.07$\pm$0.70 & $-$67.6 & 8.0 & O, IR; Late M dwarf \\%odd. g0.0+11.5 8003\\
17:11:23.02 &  $-$24:38:49.6 &  359.37 &8.74 & 0.16$\pm$0.02 & -0.64$\pm$0.64 & $-$42.8 & 9.6 & O,IR: YSO \\%g0.0+8.5 14105\\
17:12:08.72 &  $-$23:09:49.8 &  0.7 &9.45 & 0.26$\pm$0.01 & 0.16$\pm$0.31 & 20.4 & 7.8 & O,IR: YSO\\%g0.0+8.5 4104\\
17:13:34.15 &  $-$24:10:44.5 &  0.05 &8.6 & 3.88$\pm$0.01 & 0.07$\pm$0.03 & $-$1.1 & 5.6 & \\%g0.0+8.5 9432\\
17:16:29.81 &  $-$26:55:16.7 &  358.16 &6.49 & 0.19$\pm$0.01 & 0.43$\pm$0.96 & 60.6 & 12.5 & X,O,IR: YSO\\%g357+5.5 3106\\
17:17:41.97 &  $-$28:56:19.4 &  356.65 &5.11 & 0.08$\pm$0.01 & -1.12$\pm$0.96  & 64.4 & 6.3 & X,O,IR: YSO\\%g357+5.5 11755\\
17:19:56.71 &  $-$24:57:04.2 &  0.24 &6.97 & 0.32$\pm$0.01 & 1.12$\pm$0.4 & $-$27.4 & 11.1 & O,IR: YSO\\%g0.0+5.5 10507\\
17:20:18.40 &  $-$26:52:07.8 &  358.69 &5.82 & 0.20$\pm$0.01 & $-$1.87$\pm$0.31 & $-$27.8 & 7.9 & \\%g0.0+5.5 22821\\
17:20:55.04 &  $-$27:20:41.5 &  358.38 &5.44 & 0.27$\pm$0.01 & 0.15$\pm$0.31 & $-$33.8 & 12.6 & X,O,IR: Binary star\\%g357+5.5 1461\\
17:23:29.79 &  $-$28:20:36.1 &  357.87 &4.4 & 0.59$\pm$0.01 & $-$1.52$\pm$0.11 & $-$12.5 & 6.2 & \\%g357+5.5 5102\\
17:24:16.36 &  $-$27:30:43.8 &  358.66 &4.73 & 0.53$\pm$0.01 & 0.88$\pm$0.17 & 17.3 & 11.7 & X,O,IR: YSO\\%g0.0+5.5 23217\\
17:27:28.51 &  $-$25:28:20.6 &  0.76 &5.26 & 0.57$\pm$0.01 & 0.51$\pm$0.13 & 13.9 & 10.0 & O,IR: YSO\\%g0.0+5.5 6222\\
17:27:52.07 &  $-$27:27:24.6 &  359.15 &4.09 & 0.18$\pm$0.02 & $-$1.99$\pm$0.45 & 26.4 & 6.2 & \\%g0.0+5.5 19030\\
17:28:20.26 &  $-$26:08:16.6 &  0.31 &4.73 & 0.22$\pm$0.01 & $-$2.9$\pm$0.26 & 17.8 & 5.5 & \\%g0.0+5.5 9920\\
17:29:31.96 &  $-$27:31:37.8 &  359.29 &3.75 & 0.11$\pm$0.02 & $-$1.05$\pm$0.67 & 60.1 & 7.8 & \\%g0.0+2.5 13785\\
17:32:50.03 &  $-$23:54:14.4 &  2.75 &5.1 & 5.31$\pm$0.01 & $-$1.16$\pm$0.01 & $-$1.4 & 9.8 & \\%g3.0+5.5 6502\\
17:37:57.25 &  $-$27:33:05.3 &  0.29 &2.16 & 3.74$\pm$0.01 & $-$0.13$\pm$0.02 & $-$1.7 & 7.0 & \\%g0.0+2.5 7977\\
17:40:06.78 &  $-$28:05:33.9 &  0.08 &1.47 & 0.86$\pm$0.02 & $-$1.46$\pm$0.15 & 30.0 & 22.2 & \\%g0.0+2.5 9188\\
17:44:08.71 &  $-$24:24:53.4 &  3.69 &2.64 & 1.19$\pm$0.02 & $-$2.20$\pm$0.09 & 19.9 & 18.8 & \\%g3.0+2.5 1311\\
17:45:43.03 &  $-$26:58:53.9 &  1.68 &1.0 & 3.34$\pm$0.02 & $-$0.57$\pm$0.04 & $-$1.8 & 5.1 & X\\%g3.0+2.5 16065\\
17:47:18.05 &  $-$33:09:17.4 &  356.58 &$-$2.5 & 0.91$\pm$0.01 & $-$1.99$\pm$0.08 & $-$23.9 & 24.4 & \\%g357$-$2.5 12039\\
17:51:06.84 &  $-$32:18:28.5 &  357.72 &$-$2.75 & 0.58$\pm$0.01 & 0.92$\pm$0.15 & 18.4 & 14.6 & X,O,IR: YSO\\%g357$-$2.5 5244\\
17:51:15.97 &  $-$29:43:52.0 &  359.95 &$-$1.46 & 0.8$\pm$0.03 & $-$1.74$\pm$0.22 & 8.3 & 5.6 & \\%g0.0$-$2.5 10546\\
17:52:29.06 &  $-$28:49:05.6 &  0.87 &$-$1.23 & 1.07$\pm$0.02 & $-$0.86$\pm$0.12 & 9.0 & 7.1 & \\%g0.0$-$2.5 4908\\
17:53:06.42 &  $-$29:30:00.8 &  0.36 &$-$1.69 & 0.26$\pm$0.02 & $-$1.51$\pm$0.51 & 26.0 & 6.5 & \\%g0.0$-$2.5 8146\\
% next one is likely PSR J1753$-$28$ and gamma-ray source.
17:53:06.50 &  $-$28:51:26.2 &  0.91 &$-$1.37 & 0.22$\pm$0.02 & $-$0.84$\pm$0.53 & 46.0 & 8.4 & G: PSR?\\%g0.0$-$2.5 4654\\
17:53:39.01 &  $-$33:31:28.4 &  356.94 &$-$3.83 & 0.15$\pm$0.01 & $-$1.17$\pm$0.38 & 23.7 & 5.4 & \\%g357$-$2.5 9866\\
17:55:24.86 &  $-$33:58:30.7 &  356.73 &$-$4.37 & 0.16$\pm$0.01 & $-$0.28$\pm$0.37 & 39.5 & 9.0 & O,IR\\%g357$-$5.5 17737\\
17:59:20.04 &  $-$28:29:56.6 &  1.91 &$-$2.37 & 0.13$\pm$0.01 & $-$2.24$\pm$0.57 & $-$33.7 & 5.3 & \\%g3.0$-$2.5 9454\\
17:59:20.59 &  $-$30:15:48.7 &  0.37 &$-$3.24 & 0.17$\pm$0.01 & 1.66$\pm$0.58 & 22.1 & 5.2 & O,IR: $\beta$ Cep.\\%g0.0$-$2.5 8043\\
18:00:12.33 &  $-$27:44:32.5 &  2.66 &$-$2.16 & 0.12$\pm$0.02 & 0.25$\pm$1.63 & 69.1 & 8.3 & crowded field\\%g3.0$-$2.5 5429\\
18:00:32.14 &  $-$27:35:36.0 &  2.83 &$-$2.15 & 0.29$\pm$0.02 & $-$2.93$\pm$0.28 & $-$20.9 & 6.2 & \\%g3.0$-$2.5 4740\\
% Known pulsar. 18:03:35.11 &  $-$30:02:00.8 &  1.03 &$-$3.93 & 0.66$\pm$0.01 & $-$2.15$\pm$0.09 & $-$6.9 & 5.8 & g0.0$-$2.5 3845\\
18:04:03.38 &  $-$33:31:53.0 &  358.01 &$-$5.72 & 3.73$\pm$0.01 & 0.26$\pm$0.02 & $-$1.0 & 5.9 & X\\%g357$-$5.5 5387\\
% We don;t think this is a real source anymore
%18:04:33.70 &  $-$32:04:52.5 &  359.33 &$-$5.11 & 0.06$\pm$0.01 & \omit & $-$75.1 & 5.6 & \\%g0.0$-$5.5 21770\\
%Duplicate. 18:05:26.42 &  $-$29:29:54.7 &  1.69 &$-$4.02 & 0.15$\pm$0.01 & $-$4.25$\pm$0.38 & 93.1 & 19.1 & g3.0$-$2.5 10931\\
%18:05:26.45 &  $-$29:29:54.5  - this is the mosaic position. We used fit to original image. March 14/24 email.
18:05:26.44 & $-$29:29:54.6  &  1.69 &$-$4.02 & 0.18$\pm$0.01 & $-$4.31$\pm$0.38 & 86.3 & 21.5 & X,O,IR: K star\\%g3.0$-$5.5 18769\\
18:06:31.55 &  $-$29:21:37.0 &  1.93 &$-$4.16 & 0.06$\pm$0.01 & 0.27$\pm$0.99 & 85.2 & 7.0 & \\%g3.0$-$5.5 16619\\
18:11:37.21 &  $-$32:06:50.9 &  0.02 &$-$6.44 & 0.36$\pm$0.01 & $-$0.49$\pm$0.13 & 17.4 & 9.3 & \\%g0.0$-$5.5 14643\\
18:13:45.72 &  $-$33:17:38.6 &  359.18 &$-$7.39 & 0.18$\pm$0.01 & $-$1.71$\pm$0.45 & 106.8 & 24.8 & O; M dwarf\\%g0.0$-$8.5 17682\\
18:15:21.88 &  $-$32:53:02.1 &  359.71 &$-$7.49 & 3.17$\pm$0.01 & $-$0.7$\pm$0.02 & 1.3 & 5.9 & \\%g0.0$-$8.5 12510\\
18:16:34.34 &  $-$32:34:12.4 &  0.11 &$-$7.57 & 0.45$\pm$0.01 & $-$1.06$\pm$0.11 & 7.2 & 5.1 & \\%g0.0$-$8.5 9003\\
18:17:09.82 &  $-$33:17:31.2 &  359.52 &$-$8.02 & 0.13$\pm$0.01 & 0.62$\pm$0.48 & $-$24.6 & 5.5 & O,IR: YSO\\%g0.0$-$8.5 14366\\
18:18:31.06 &  $-$32:54:18.2 &  360.0 &$-$8.09 & 0.2$\pm$0.01 & $-$1.18$\pm$0.25 & 45.7 & 14.2 & \\%g0.0$-$8.5 9917\\
18:19:52.24 &  $-$29:16:34.4 &  3.38 &$-$6.68 & 0.27$\pm$0.02 & 0.68$\pm$0.63 & $-$27.2 & 5.1 & O,IR: Binary star\\%g3.0$-$5.5 3249\\
18:29:52.06 &  $-$33:55:40.8 &  0.13 &$-$10.67 & 0.2$\pm$0.01 & 1.2$\pm$0.4 & 41.2 & 12.2 & O,IR: YSO\\%g0.0$-$11.5 11597\\
18:29:52.97 &  $-$33:43:04.7 &  0.33 &$-$10.58 & 0.16$\pm$0.01 & 0.61$\pm$0.37 & 46.0 & 11.0 & O,IR: Star\\%g0.0$-$11.5 9720\\
18:34:49.11 &  $-$35:05:25.8 &  359.51 &$-$12.09 & 5.4$\pm$0.01 & $-$0.5$\pm$0.02 & $-$1.2 & 9.6 & \\%g0.0$-$11.5 16490\\
18:35:01.56 &  $-$34:18:20.1 &  0.25 &$-$11.79 & 3.18$\pm$0.01 & 0.33$\pm$0.02 & 1.1 & 5.1 & \\%g0.0$-$11.5 10505\\
18:41:54.78 &  $-$36:07:49.2 &  359.14 &$-$13.83 & 0.14$\pm$0.01 & $-$1.58$\pm$0.55 & $-$38.4 & 8.5 & \\%g0.0$-$14.5 23097\\
18:43:58.38 &  $-$35:59:10.0 &  359.45 &$-$14.15 & 0.28$\pm$0.01 & 1.31$\pm$0.27 & 12.3 & 5.7 & O,IR: YSO\\%g0.0$-$14.5 20286\\
18:50:19.66 &  $-$36:29:12.8 &  359.49 &$-$15.53 & 3.97$\pm$0.01 & $-$0.14$\pm$0.01 & 1.0 & 6.5 & \\%g0.0$-$14.5 19904\\
18:52:26.04 &  $-$37:30:37.8 &  358.67 &$-$16.3 & 0.21$\pm$0.01 & $-$2.70$\pm$0.37 & 41.2 & 9.5 & X,O,IR: M4.5 dwarf\\ %X,O,IR: L\,489-43 (odd) %g0.0$-$17.5 25349\\
18:57:45.15 &  $-$37:19:38.3 &  359.25 &$-$17.22 & 0.53$\pm$0.01 & $-$0.05$\pm$0.12 & $-$28.7 & 22.7 & X,O,IR: YSO\\%g0.0$-$17.5 20666\\
19:00:51.24 &  $-$36:13:15.8 &  0.57 &$-$17.39 & 0.06$\pm$0.01 & $-$0.63$\pm$0.69 & 53.4 & 5.2 & IR: YSO \\%g0.0$-$17.5 8226\\
19:01:40.56 &  $-$36:44:32.4 &  0.12 &$-$17.74 & 0.08$\pm$0.01 & 2.08$\pm$0.76 & 63.3 & 9.4 & O,IR: YSO\\%g0.0$-$17.5 12658\\
19:01:48.04 &  $-$36:57:20.5 &  359.92 &$-$17.84 & 0.61$\pm$0.01 & 0.56$\pm$0.12 & 18.0 & 18.7 & IR: YSO\\%g0.0$-$17.5 14425\\
19:01:55.64 &  $-$37:39:41.1 &  359.24 &$-$18.11 & 5.34$\pm$0.01 & 0.75$\pm$0.01 & $-$1.5 & 13.0 & X,O,IR: YSO\\%g0.0$-$17.5 20715\\
19:03:06.35 &  $-$37:16:41.7 &  359.7 &$-$18.19 & 0.16$\pm$0.01 & $-$0.55$\pm$0.28 & 44.8 & 12.2 & O,IR: YSO\\%g0.0$-$17.5 16388\\
19:06:25.66 &  $-$37:03:49.5 &  0.16 &$-$18.74 & 0.19$\pm$0.01 & $-$0.18$\pm$0.32 & 85.6 & 24.7 & \omit \\% Gam\,CrA A (odd) g0.0$-$17.5 12281\\
19:07:38.78 &  $-$37:08:55.9 &  0.16 &$-$19.0 & 6.64$\pm$0.01 & $-$1.08$\pm$0.01 & $-$1.1 & 7.7 & \\ %g0.0$-$17.5 12238\\
\enddata
\tablecomments{O=Optical counterpart, IR=infrared counterpart, X=X-ray counterpart and G=gamma-ray counterpart. YSO=young stellar object. Details on individual source detections are given in \S\ref{sec:count}}
\end{deluxetable}

\subsection{Linearly Polarized Sources}\label{sec:lin-candi}

A different approach was needed for identifying linear polarized pulsar candidates. The Stokes Q and U mosaic images have rich structure, likely due to Faraday rotation from the interstellar medium of large scale Galactic emission. Moreover, the areal density of linearly polarized sources in the master catalog is large, approximately 40 deg$^{-2}$, with the majority of the polarized emission coming from extended sources (AGN jets, supernova remnants, etc.). More stringent selection criteria were needed to achieve a manageable number of polarized sources. For each mosaic pointing we first eliminated extended sources by requiring that the compactness criterion exceed 0.9 and that the source not be fit by multiple Gaussian components (see \S\ref{sec:catalog}). 

Once we identified a sample of compact radio sources, we further required that the spectral index $\alpha\lesssim -1.5$ and the fractional linear polarization $\vert{\rm{P/I}}\vert>5\%$. These criteria restrict the phase space to pulsar-like properties in order to keep the false positives to a manageable level, but also introduces a selection bias in the sample. Approximately 50\% of all known pulsars have spectral indices steeper than $-1.5$ \citep{2018MNRAS.473.4436J} and 95\% of pulsars have linear polarization above 5\% \citep{2018MNRAS.474.4629J}. Despite these more restrictive criteria, visual inspection of the polarization images resulted in about 80\% of the detections being identified as either image artifacts or extra-galactic. The largest false positives were extragalactic sources, seen as compact polarized knots in jets, double radio sources, and clusters of radio sources. Our final sample of 10 linear polarized sources is given in Table \ref{tab:linear} and is plotted in Fig. \ref{fig:polar}. The "P" entries in column 6 of Table \ref{tab:point} lists the number of linear polarized sources found in each of the mosaic pointings, including two sources identified in more than one mosaic field.

The properties of these linear polarized sources are listed in Table \ref{tab:linear} and plotted in Figure \ref{fig:polar} (right). Not surprisingly, given our selection criteria, the spectral index distribution and high fractional polarization are similar to the known pulsars in Table \ref{tab:known} and the larger sample of linear polarization measurements from \citet{2018MNRAS.474.4629J}. There are some compelling pulsar candidates in Table \ref{tab:linear} based on the compactness, steep spectrum and polarization criteria. J173121.50$-$250235.7 is an exceptionally bright steep spectrum source with a Stokes I flux density of 12 mJy and is 16\% polarized, while J174408.71$-$242453.4 is also bright (I=1.2 mJy) and steep spectrum but has both significant linear {\it and} circular polarization. It is likely, however, that these sources are relatively nearby and not in the bulge. Given our central observing frequency ($\nu$=1284 MHz) and the bandwidth of the 12 sub-band averages ($\delta\nu\simeq$50 MHz) that we used (\S\ref{sec:obs}), we expect significant Faraday bandwidth depolarization for rotation measures (RM) in excess of 235 rad m$^{-2}$ \citep[see Eqn 1 of ][]{2021MNRAS.507.3888H}, well below RM values reported near the GC \citep{2021MNRAS.502.3814L}.

Our main conclusion from this pilot effort is that linear polarized images are a sub-optimal method for identifying pulsar candidates. While we are confident that the detections in Table \ref{tab:linear} are real linearly polarized sources in the images, the overwhelming number of artifacts and other real source populations required additional selection criteria that introduce significant selection biases making the method ill suited for surveys of the bulge or elsewhere in the Galaxy.

 %This hypothesis is further supported by the source distribution in Figure \ref{fig:polar}. The mean spectral index of our 61 candidates is $-0.6$, versus $-1.7$ for the known pulsars in Table \ref{tab:known}. The fractional polarization for the known pulsars ranges from 2-20\%, while the candidates are distributed over a much larger range. The candidates at higher polarization values could still be pulsar candidates because of a noise bias at low flux densities (REF). The absence of sources in the left hand corner of this plot is a result of lower S/N cutoff, but the lack of bright, highly polarized sources is real. As we noted above the cluster of flat-spectrum, weakly polarized (1-2\%) could have some false positives among them due to beam squint effects. In \S\ref{sec:count} we will use multi-wavelength counterparts in an effort to distinguish promising pulsar candidates from other source populations. 
 
\begin{deluxetable}{llllrrrr}
\tablecaption{Linear Polarized Sources \label{tab:linear}}
\tablehead{\colhead{RA} & \colhead{DEC} & \colhead{$l$} & \colhead{$b$} &  \colhead{I} & \colhead{$\alpha$} & \colhead{P/I} &  \colhead{S/N} \\
\colhead{(h~~m~~s)} & \colhead{($^\circ$\ ~~$'$\ ~~$"$)} & \colhead{(deg)} & \colhead{(deg)} & \colhead{(mJy)} & \colhead{} & \colhead{(\%)}  & \colhead{}
}
\startdata
17:01:54.67 &  $-$22:50:08.3 &  359.58 & 11.55 & 0.73$\pm$0.01 & $-$2.00$\pm$0.09 & 13.3 & 11.6 \\%& g0.0+11.5 10516\\
17:18:37.15 &  $-$23:35:10.5 &  1.21 &7.99 & 1.62$\pm$0.02 & $-$1.52$\pm$0.11 & 7.2 & 10.4 \\%& g0.0+8.5 939\\
17:26:26.80 &  $-$30:48:09.2 &  356.19 &2.5 & 3.63$\pm$0.03 & $-$1.65$\pm$0.08 & 6.9 & 14.9 \\%& g357+2.5 13725\\
17:31:21.50 &  $-$25:02:35.7 &  1.6 &4.76 & 12.33$\pm$0.02 & $-$1.48$\pm$0.01 & 14.0 & 197.0 \\%& g3.0+5.5 14570\\
%Known pulsar. 17:34:41.37 &  $-$24:16:15.0 &  2.67 &4.54 & 0.6$\pm$0.01 & $-$1.45$\pm$0.1 & 46.8 & 30.2 & g3.0+5.5 6982\\
%Known pulsar (RRAT) 17:39:32.79 &  $-$25:21:09.3 &  2.34 &3.03 & 0.35$\pm$0.01 & $-$2.54$\pm$0.16 & 15.9 & 6.2 & g3.0+2.5 11384\\
17:44:08.71 &  $-$24:24:53.4 &  3.69 &2.64 & 1.19$\pm$0.02 & $-$2.20$\pm$0.09 & 14.3 & 12.8 \\%& g3.0+2.5 1311\\
18:18:22.84 &  $-$32:44:16.0 &  0.13 &$-$7.99 & 0.23$\pm$0.01 & $-$2.34$\pm$0.26 & 31.0 & 12.1 \\%& g0.0$-$8.5 8790\\
18:21:46.77 &  $-$32:47:27.8 &  0.41 &$-$8.65 & 0.26$\pm$0.01 & $-$1.48$\pm$0.33 & 16.6 & 7.3 \\%& g0.0$-$8.5 6574\\
18:24:17.96 &  $-$32:38:05.4 &  0.79 &$-$9.05 & 0.60$\pm$0.06 & $-$2.66$\pm$0.54 & 23.7 & 13.6 \\%& g0.0$-$8.5 3995\\
18:27:31.72 &  $-$32:52:38.9 &  0.88 &$-$9.77 & 1.33$\pm$0.02 & $-$1.51$\pm$0.08 & 5.9 & 12.3 \\%& g0.0$-$8.5 3455\\
18:37:36.28 &  $-$33:50:13.8 &  0.92 &$-$12.08 & 0.25$\pm$0.01 & $-$3.04$\pm$0.29 & 16.3 & 7.4 \\%& g0.0$-$11.5 4028\\
\enddata
%\tablecomments{}
\end{deluxetable}

 \subsection{Low-Band Circularly Polarized Sources}\label{sec:lowhangingcandi}

There are significant numbers of sources in the catalog that are not detected with 5$\sigma$ significance across the {\it full} frequency range of the MeerKAT L-band receivers. This may be the result of channel flagging due to RFI, or that the source has a sufficiently steep spectrum that it falls below the noise threshold at higher frequencies. As we are looking for steep-spectrum pulsars, we considered additional sources which met our selection criteria using only the {\it lower} half of the band centered at 1.022 GHz. We might expect that the majority of such sources will be weak and steep spectrum in order for them to have been missed by the selection criteria in \S\ref{sec:candi} and \S\ref{sec:lin-candi}. 

Using only the low frequency end of L-band, we found a total of 9 compact, circularly polarized ($\vert$V/I$\vert>$1\%) sources and 32 linearly polarized (P/I$>$5\%) sources. None of the low-band linear polarized sources passed our visual inspection. Of the 9 low-band circularly polarized sources, three were image artifacts and one was an known pulsar, PSR\,J1748$-$3009 (Table \ref{tab:known}). The final list of five sources is shown in Table \ref{tab:lowbandcirc}. The full-band peak flux density is listed, whereas the percent polarization and S/N are derived from the low end of the band. We note that one of these sources, J182417.96$-$323805.4, is also detected as a linearly polarized source in Table \ref{tab:linear}. 

\begin{deluxetable}{llllrrrr}
\tablecaption{Low-Band Circularly Polarized Sources \label{tab:lowbandcirc}}
\tablehead{\colhead{RA} & \colhead{DEC} & \colhead{$l$} & \colhead{$b$} &  \colhead{I} & \colhead{$\alpha$} & \colhead{V/I} &  \colhead{S/N} \\
\colhead{(h~~m~~s)} & \colhead{($^\circ$\ ~~$'$\ ~~$"$)} & \colhead{(deg)} & \colhead{(deg)} & \colhead{(mJy)} & \colhead{} & \colhead{(\%)}  & \colhead{}
}
\startdata
17:27:34.51 &  $-$26:09:57.1 & 0.19 & 4.86 & 0.07$\pm$0.01 & $-$1.87$\pm$0.72 & 64.0 & 5.5 \\%& g0.0+5.5 10946\\
17:31:14.95 &  $-$27:35:32.4 & 359.45 & 3.39 & 0.30$\pm$0.01 & $-$2.18$\pm$0.22 & 13.9 & 5.3 \\%& g0.0+2.5 12921\\
18:02:21.53 & $-$32:54:38.0 & 358.38 & $-$5.10 & 3.46$\pm$0.01 & $-$1.61$\pm$0.01 & 1.3 & 5.5 \\%& g357.0-5.5 1740\\
% This one is also in Table 4 (linearly polarized):
18:24:17.96 &  $-$32:38:05.4 & 0.79 & $-$9.05 & 0.60$\pm$0.06 & $-$2.66$\pm$0.54 & 12.4 & 6.2 \\%& g0.0-8.5 3995\\
18:51:20.86 &  $-$35:46:19.5 &  0.26 & $-$15.44 & 0.18$\pm$0.01 & $-$1.36$\pm$0.29 & 20.0 & 5.3 \\%& g0.0-14.5 12466\\
\enddata
%\tablecomments{}
\end{deluxetable}

\section{Additional Selection Criteria}\label{sec:count}

As previously noted in \S\ref{sec:candi}, there are other source populations among the circularly polarized sources in Table \ref{tab:circ} that are not pulsars. This is less likely for the linear polarized sources in Table \ref{tab:linear} since we have used more stringent pulsar-like criteria to limit the false positives. Extragalactic radio sources are not expected to be a significant contaminant for circular polarization. Apart from fast radio bursts, which we do not expect to detect, the circular polarization from extragalactic radio sources rarely exceeds 1\% \citep{2002PASA...19...43M}. On the other hand, magnetically active single and binary stars can produce non-thermal (coherent and incoherent) radio emission that can be strongly polarized. A comprehensive list of the types of radio stars expected and the origin of their polarized emission can be found in 
\cite{2021MNRAS.502.5438P} and \cite{2023A&A...670A.124C} for centimeter and decimeter wavelengths, respectively.

A multi-wavelength approach can be useful for both identifying these stellar populations and for bolstering pulsar candidates. To this end, in the next several subsections we search for counterparts to our polarized radio sources at gamma-rays, X-rays, optical and infrared wavelengths. Where possible, we also include archival radio data to better constrain the in-band spectral indices and to search for variability. However, before focusing on any wavelength-specific searches, we began by searching a 2-arcmin radius around 
each of the sources in Tables \ref{tab:circ}-\ref{tab:lowbandcirc} using the SIMBAD database \citep{2000A&AS..143....9W}. There are two immediate findings from this search that help to inform the multi-wavelength searches below. The first is that given our astrometry accuracy (\S\ref{sec:properties}), the high stellar densities in the bulge will likely result in significant false positives in the optical/NIR. We discuss our efforts to mitigate this in \S\ref{sec:oir}. The second finding is that there are a large number of unexpectedly bright, nearby stars or young stellar objects within 5-arcsec of our circularly polarized sources. For example, one of our strongly polarized sources (V/I=86\%), J190625.66$-$370349.5 lies close to the high proper motion, naked eye star $\gamma$ CrA with G=4.8 mag.  Proper motion corrections will be needed to assess whether these are real associations (see \S\ref{sec:oir}), but the preponderance of SIMBAD matches of stars with G-band magnitudes in excess of 10-15 mag is suggestive, given their rarity among the known {\it Gaia} stellar brightness distribution \citep{2021A&A...649A...1G}. 

\subsection{High Energy Counterpart Searches}\label{sec:highenergy}

\begin{figure}
\centering
\includegraphics[width=\textwidth]{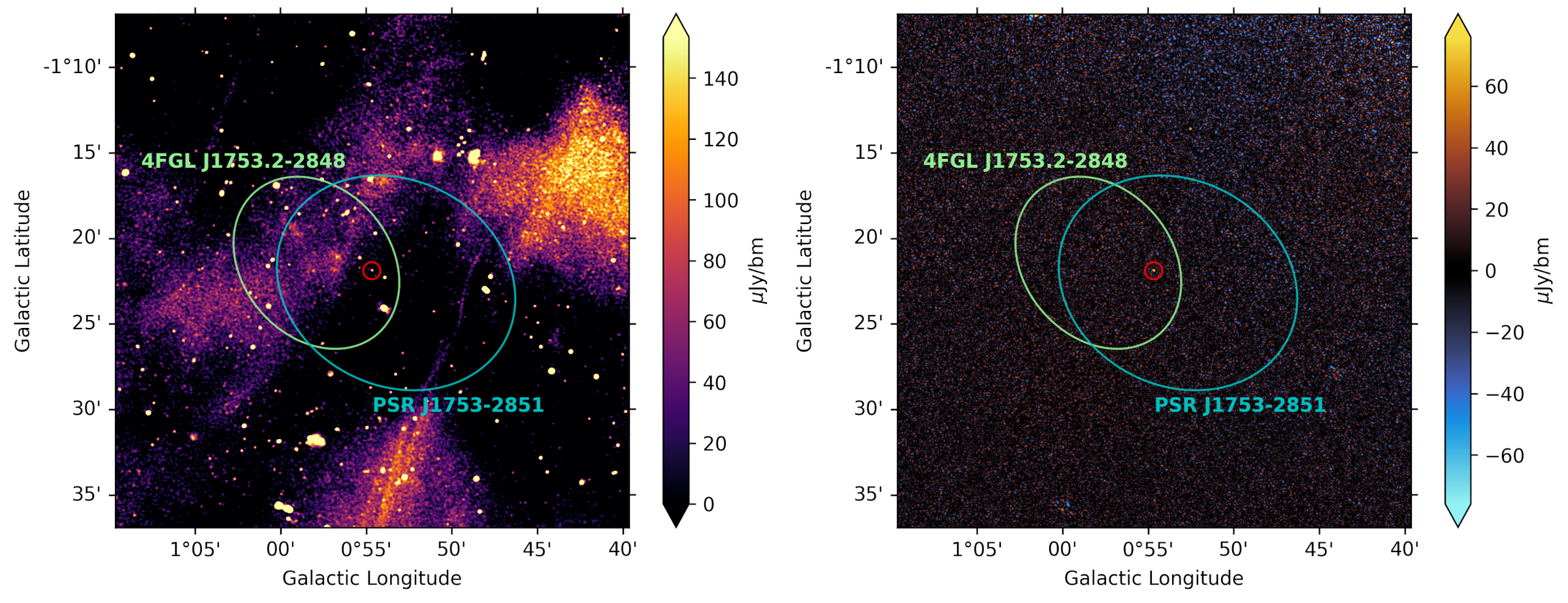}
\caption{Stokes I (left) and Stokes V (right) cutout images centered on our circularly polarized source J175306.50$-$285126.2 (red circle). We show the error circles for the {\it Fermi} source 4FGL\,J1753.2$-$2848 and the known pulsar  PSR\,J1753$-$28. A possible association is suggested but not proven between our polarized radio source, the known pulsar and the gamma-ray source.}
\label{fig:fermipsr}
\end{figure}

The {\it Fermi} Gamma-ray Space Telescope has discovered pulsed emission from nearly 300 pulsars \citep{2023ApJ...958..191S}, but there remain many more candidates with distinct spectral and temporal signatures among the unknown or unassociated gamma-ray sources~\citep{2022ApJS..260...53A}. We searched for matches between our polarized pulsar 
sources and the 2577 unknown or unassociated \emph{Fermi} gamma-ray sources in the latest 14-year {\it Fermi}-LAT incremental catalog \citep{2023arXiv230712546B}. We find one match of 4FGL\,J1753.2$-$2848 with the circularly polarized source J175306.50$-$285126.2 ($\vert{\rm{V/I}}\vert$=46\%) in Table \ref{tab:circ}. 
The unknown {\it Fermi} source is associated to the X-ray source 1RXS J175328.7$-$285014, and 
its best-fit energy spectrum is compatible with pulsar spectral characteristics~\citep{2023ApJ...958..191S}.
We note that the 86 msec period PSR\,J1753$-$28 lies 94-arcsec from this polarized radio source, but the pulsar position is poorly constrained \citep[$\pm{7}^\prime$;][]{2020MNRAS.493.1063C}, preventing us from making a positive identification (see Fig.\ref{fig:fermipsr}). An improved timing position for PSR\,J1753$-$28 and a measurement of its period derivative and spindown energy, could test whether this polarized radio source is the pulsar and whether it is capable of powering the gamma-ray emission. None of the 
sources in Tables \ref{tab:linear} and \ref{tab:lowbandcirc} have \emph{Fermi} counterparts.

Young radio pulsars and MSPs show a scaling of their spindown energy ($\dot{\rm{E}}$) with X-ray flux \citep[e.g.,][]{2002A&A...387..993P}. To this end we searched for X-ray counterparts of our polarized radio sources (Tables \ref{tab:circ}-\ref{tab:lowbandcirc}) in the \emph{Chandra} and \emph{SWIFT} bulge surveys \citep{2011ApJS..194...18J,2014ApJS..210...18J,2021MNRAS.501.2790B}. No matches were found at the lower Galactic latitudes where these X-rays surveys were carried out. 
There was more success using the Chandra Source Catalog 2.0  \citep{2010ApJS..189...37E} and the recently released 6-month sky survey with the eROSITA telescope array in the 0.2-2.3 keV energy range \citep{2024A&A...682A..34M}. In the Chandra catalog we found two point sources within 4$^{\prime\prime}$ of the circularly polarized sources in Table \ref{tab:circ}. In the eRosita catalog we found eight X-ray sources that lie within the joint radio/X-ray position uncertainties, despite the fact that this initial data release covers only half of our bulge survey area (i.e., all Galactic latitudes for  longitudes west of Sgr A*). There were no X-ray counterparts to any of the polarized sources in Tables \ref{tab:linear} and \ref{tab:lowbandcirc}.

Most of the radio spectra of these X-ray sources are flat or inverted (i.e., $\alpha\gtrsim -0.6$) with one exception. J180526.44$-$292954.6 has $\alpha=-4.31$ and appears to be associated with 2CXO\,J180526.4-292952. We discuss this and other sources in more detail in \S\ref{sec:oir} and we show that the majority of the X-ray associations likely originate from active stars and young stellar objects. All of these gamma-ray and X-ray matches are identified in the Table \ref{tab:circ} as G or X, respectively.

\subsection{Optical and Infrared Counterparts}\label{sec:oir}

Our search for optical and infrared counterparts of the polarized sources (Tables \ref{tab:circ}-\ref{tab:lowbandcirc}) used the latest {\it Gaia} data release \citep{2023A&A...674A...1G} and the ALLWISE catalog from the Wide-field Infrared Survey Explorer \citep{2010AJ....140.1868W}. Astropy code was written in order to make proper motion corrections of the {\it Gaia} stars to the mean epoch of the MeerKAT observations (2020.3) for all radio sources in Tables \ref{tab:circ}-\ref{tab:lowbandcirc}.

Most isolated pulsars do not have detectable O/IR emission \citep{2021MNRAS.501.1116A}.  Thus the detection of optical/IR emission at centimeter wavelengths generally means that the circularly polarized source is a star, e.g., magnetically active dwarfs, young stellar object (YSOs) and interacting binaries \citep{2021MNRAS.502.5438P}. However, caution is warranted since our main goal is identifying MSP candidates in the bulge and they can have a faint optical/IR companion. In the compilation of \cite{2023MNRAS.525.3963K} the majority of known compact MSP binaries have G-band magnitudes companions fainter than 18$^{th}$ mag, but there are a small number of systems in the 14-18 mag range. When possible we use other criteria (e.g., colors) in looking at matches in this magnitude range. 

\cite{2018MNRAS.481.2148W} point out two additional challenges. The first problem is false identifications arising due to the high star density and extreme crowding in the bulge.  For each possible association of a star of magnitude G (or W1 for WISE) we estimate the probability of chance coincidence by using the number density of stars above this magnitude within a 1-arcmin radius of the radio source. We flag as suspect any associations with probabilities above 0.5\%. In practice, given our astrometric error, this resulted in a magnitude cutoff of about G$>$15.6 mag for {\it Gaia} sources. As a final check we visually inspect available optical and WISE images of each field to identify directions with extreme crowding. A second issue raised by \cite{2018MNRAS.481.2148W} is the non-identification of the brightest sources (G$<10$ or W1$<$8 mag). This is in part due to position errors caused by the saturation of the detectors, and small systematic effects in the proper motion of stars \citep{2021A&A...649A.124C}. 

None of the 10 (steep-spectrum) linear polarized sources in Table \ref{tab:linear} had IR counterparts or {\it Gaia} proper motion corrected matches within the uncertainty of our radio positions. A similar null result was obtained for the sources in Table \ref{tab:lowbandcirc}.

The source matching of the ALLWISE catalog with our circularly polarized sources in Table \ref{tab:circ} gave an unexpected result. Most of the 28 positive matches %within a 2$^{\prime\prime}$ radius 
came from sources with a spectral index $\alpha>-0.55$. As this is nearly half our sample, this result points to a real (non-pulsar) source population. Using the WISE colors we identified that a significant fraction of these matches were likely from YSOs \citep{2014ApJ...791..131K}. Radio emission from YSOs has been detected with a range of spectral indices similar to what is seen here (i.e., $-0.6$ to +1). This emission is thought to originate via either thermal free-free emission from ionized gas or non-thermal emission generated by jet shocks. YSOs can be time-variable, with high degrees of circular polarization, and may have counterparts at X-rays, infrared and sub-millimeter wavelengths \citep{1998ApJ...494L.215F, 2006A&A...446..155F,2021A&A...653A.101F}. Further support for these YSO identifications came from SIMBAD and the {\it Gaia}/ALLWISE catalog of YSO candidates \citep{2019MNRAS.487.2522M}. All of these YSO matches are identified in Table \ref{tab:circ} as``YSO".

Apart from young stellar objects associated with star formation, there is a heterogeneous collection of radio-emitting stars. For example, J165648.02$-$230901.8 is 24 Oph.\,A (HD\,152849), a 6$^{th}$ magnitude star of spectral type A05, and may be a new case of (polarized) radio emission from a rapidly-rotating A star \citep{2021ApJ...912L...5W}. J172055.04$-$272041.5 is HD\,156848B, part of a spectroscopic binary (G=9.1 mag). According to SIMBAD, one member is a F7/8 subgiant star while the other component being either an F or G type star, and is listed as a possible ROSAT source by \cite{2009ApJS..184..138H}. 175920.59$-$301548.7 is 
HD\,316903 and is listed in SIMBAD as a pulsating variable of spectral class B8, making it only the second $\beta$ Cepheid variable for which radio emission has been detected \citep{2014RMxAA..50..127T}. J181952.24$-$291634.4 is HD\,168210, a bright (G=8.7 mag) eclipsing binary dominated by a G5 main sequence star. It is listed as a possible member of the nearby $\beta$ Pictoris moving group, and is a likely ROSAT source \citep{2009ApJS..184..138H,2018ApJ...856...23G,2023ApJ...946....6C}.

There are at least three nearby M dwarfs in our sample, which appear to be fully convective according to their {\t Gaia} colors and absolute magnitudes. J170027.74$-$220742.7 is the high proper motion star, 2MASS J17002789$-$2207322. It has the colors of a late M dwarf \citep{2013ApJS..208....9P}\footnote{\url{https://www.pas.rochester.edu/~emamajek/EEM_dwarf_UBVIJHK_colors_Teff.txt}} and it is listed as an ultracool dwarf in the planet transit candidate survey by \citet{2021A&A...645A.100S}. J185226.04$-$373037.8 is our most polarized source in the sample. J181345.72$-$331738.6 has been previously cataloged in SIMBAD as L\,489-43, a high proper motion M4.5 dwarf with X-ray emission \citep{2006AJ....132..866R}.

Our most unusual source is J180526.44$-$292954.6, with extreme polarization (V/I=86.3\%) and spectral index $\alpha=-4.31$ values. The radio source is coincident with HD\,317101A, the bright primary (G=9.9 mag) of a nearby (d=33 pc) high proper motion (146.5 mas/yr) visual binary system \citep{2021AJ....161..147B}. Based on the temperature and colors of these stars, both appear to be K dwarfs \citep{2012ApJ...746..154P,2019AJ....158..138S}. As noted in \S\ref{sec:highenergy}, HD\,317101A is also a X-ray source. The radio source was observed in several MeerKAT pointings, three at or within the half-power point of the primary beam on 2020 June 28 and July 10, and each consisting of 12 five-min scans observed over 9 hours. Images made from the combination of all 12 scans on each date vary by $\sim$25\% from each other. 

With its strong variability, an active phase that persists for weeks or months, a high degree of polarization and a very steep radio spectrum, J180526.44$-$292954.6 appears to share many of the characteristics of a group of GC radio transients \citep[GCRT;][]{2005Natur.434...50H, 2007ApJ...660L.121H, 2009ApJ...696..280H,2021ApJ...920...45W}. However, unlike other GCRTs \citep[e.g.,][]{1979ApJ...228..755R,2008ApJ...687..262K}, we appear to have identified a likely multi-wavelength counterpart and its distance. While time-variable, polarized radio (and X-ray) emission has been seen toward K dwarfs \citep{1992A&A...264L..31G,2021MNRAS.502.5438P}, the properties of this source seem extreme and are thus worthy of further study.

\subsection{Radio Variability}\label{sec:var}

In addition to pulse-to-pulse fluctuations, the flux density of pulsars can vary significantly on longer timescales due to intrinsic effects (e.g. nulling, eclipses) and extrinsic effects (e.g., refractive and diffractive scattering). While variability was used as an early pulsar search strategy \citep[aka "scintars";][]{1979ApJ...228..755R}, and it helped to identify the first millisecond pulsar \citep{1982Natur.300..615B}, it has only recently been used in conjunction with other search criteria in imaging surveys \citep[i.e.,][]{2021ApJ...920...45W,2023PASA...40....3S}. 

\begin{figure}[htb]
    \centering
    \includegraphics[width=3in]{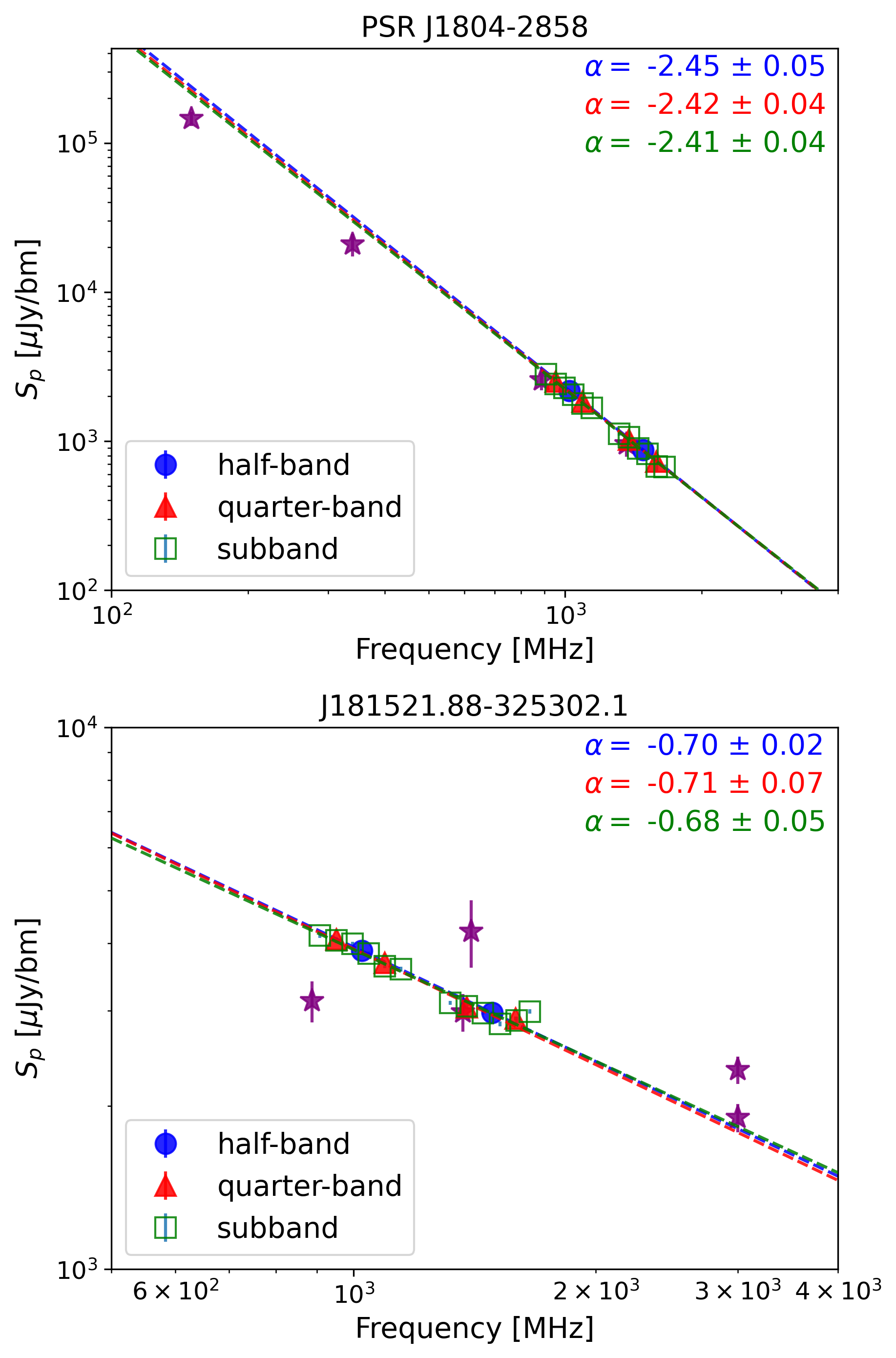}
    \caption{{\it Top}: Spectrum of pulsar J1804-2858. \textbf{Purple stars} plot peak intensities from the TGSS \citep[150 MHz,][]{2017A&A...598A..78I}{}, VLITE \citep[340 MHz,][]{2016SPIE.9906E..5BC}{}{}, RACS-low \citep[888 MHz,][]{2021PASA...38...58H}{}, and RACS-mid \citep[1368 MHz,][]{2023PASA...40...34D}{}{} catalogs. \textbf{Blue circles, red triangles, and green squares} show flux measurements from the MeerKAT half-band, quarter-band, and subband images. The corresponding in-band spectral fits (dashed lines) show excellent agreement both in-band and with the other radio catalogs. {\it Bottom}: Spectrum of the polarized source J181521.88-325302.1 exhibiting variability. Catalog values are from RACS-low, RACS-mid, NVSS \citep[1.4 GHz,][]{1998AJ....115.1693C}{}{}, and the first two VLASS epochs \citep[3 GHz,][]{2020PASP..132c5001L}{}{}.} \label{spectra}
\end{figure}

Our approach to studying variability was to look in existing archival surveys for radio flux density measurements of the pulsar candidates in Tables \ref{tab:circ}-\ref{tab:lowbandcirc}. We began first by excluding those 28 sources previously identified as stars in \S\ref{sec:oir}, resulting in 47 candidates. We next looked at existing archival datasests or surveys which were observed at or near the same frequency of this MeerKAT bulge dataset \citep{1998AJ....115.1693C,2021PASA...38...54M,2021PASA...38...58H,2024PASA...41....3D}. Recall that our MeerKAT flux densities in Tables \ref{tab:circ}-\ref{tab:lowbandcirc} were derived from the mosaic images.  Although this results in fitted MeerKAT spectra with very low uncertainties, they are not useful for studying variability alone since they are time-averages over multiple pointings. We consider a L-band source to be a variable source if a survey flux density differs from the MeerKAT interpolated value by more than 20\% and 3$\sigma$. This constraint yields 11 sources with 25 - 60\% variations based on available survey results. The circularly polarized sources J181521.88$-$325302.1 (shown in Fig. \ref{spectra}) and J172329.79$-$282036.1 are notable examples of variability. The latter is detected at 887 MHz with a flux density 60\% and 4$\sigma$ higher than expected from the MeerKAT spectrum. Also of note is an additional tenth source, J180032.14$-$273536.0, which has no survey detections, but a 5$\sigma$ upper limit at 887 MHz that is 50\% lower than the expected value. We include the spectrum of PSR\,J1804$-$2858 in Fig. \ref{spectra} to illustrate the consistency between the MeerKAT in-band measurements of the spectral index, extrapolated to archival survey measurements.

A second, less reliable, measure of variability comes from comparing our results with surveys at different sky frequencies \citep{2017A&A...598A..78I,2017MNRAS.464.1146H,2019PASA...36...47H,2020PASP..132c5001L}. In addition to these published surveys, we inspected the extensive database from the VLA Low-band Ionosphere and Transient Experiment (VLITE) which regular monitors the radio sky at 340 MHz commensal with the regular observing at the VLA \citep{2016SPIE.9906E..5BC,2016ApJ...832...60P}. In all these instances we are looking for significant deviations from the power law fits extrapolated outside the MeerKAT band. Such deviations are less robust reliable indicators of variability since they could originate from real deviations in the pulsar's spectrum from a simple power-law.

We list variable radio sources in column six of Table \ref{tab:boolean}. Radio sources with archival flux densities measured near or within the MeerKAT L band are either labeled Y1 or N, indicating variable or not-variable, respectively. Note that the variability timescale is poorly constrained as the epochs vary from months to decades. Variability outside of the MeerKAT frequency range (i.e. significant deviations from power-law extrapolations) is labeled as Y2. Sources without archival radio observations are left blank. We will discuss variability and other properties in Table \ref{tab:boolean} in more detail in \S\ref{sec:results}. 

\begin{deluxetable}{lcccccc}
\tablecaption{Pulsar Candidate Properties}\label{tab:boolean}
\tablehead{\colhead{J2000 Name} & \colhead{Compact} & \colhead{Steep} &  \colhead{C Pol.} & \colhead{L Pol.} & \colhead{Variable} &  \colhead{HE}
}
\startdata
% Now thought to be a star
%170027.74$-$220742.7 & Y & N & Y & N & \omit & \omit \\%odd. g0.0+11.5 8003\\
  171334.15$-$241044.5 & Y & N & Y & N & Y1& \omit \\ %g0.0+8.5 & 9432
  {\it 172018.40$-$265207.8} & Y & Y & Y & N & \omit & N\\ %g0.0+5.5 & 22821\\
  {\it 172329.79$-$282036.1} & Y & Y & Y & N & Y1& N\\ %g357+5.5 & 5102\\
  {\it 172752.07$-$272724.6} & Y & Y & Y & N & \omit & N\\ %g0.0+5.5 & 19030\\
  {\it 172820.26$-$260816.6} & Y & Y & Y & N & N & \omit \\ %g0.0+5.5 & 9920\\
  172931.96$-$273137.8 & Y & N & Y & N & \omit & N\\ %g0.0+2.5 & 13785\\
  173250.03$-$235414.4 & Y & N & Y & N & Y1& \omit \\ %g3.0+5.5 & 6502\\
  173757.25$-$273305.3 & Y & N & Y & N & N & \omit \\ %g0.0+2.5 & 7977\\
  {\it 174006.78$-$280533.9} & Y & N & Y & N & N& \omit \\ %g0.0+2.5 & 9188\\
  {\it 174408.71$-$242453.4} & Y & Y & Y & Y & Y1 & \omit \\ %g3.0+2.5 & 1311\\
  174543.03$-$265853.9 & Y & N & Y & N & Y2& Y\\ %g3.0+2.5 & 16065\\
  {\it 174718.05$-$330917.4} & Y & Y & Y & N & N & N\\ %g357-2.5 & 12039\\
  {\it 175115.97$-$294352.0} & Y & Y & Y & N & N & \omit \\ %g0.0-2.5 & 10546\\
  {\it 175229.06$-$284905.6} & Y & N & Y & N & N & \omit \\ %g0.0-2.5 & 4908\\
  {\it 175306.42$-$293000.8} & Y & Y & Y & N & N & \omit \\ %g0.0-2.5 & 8146\\
  {\it 175306.50$-$285126.2}$^a$ & Y & N & Y & N & \omit & Y\\ %g0.0-2.5 & 4654\\
  {\it 175339.01$-$333128.4} & Y & N & Y & N & N & N\\ %g357-2.5 & 9866\\
  {\it 175920.04$-$282956.6} & Y & Y & Y & N & N & \omit \\ %g3.0-2.5 & 9454\\
  180012.33$-$274432.5 & Y & N & Y & N & \omit & \omit \\ %g3.0-2.5 & 5429\\
  {\it 180032.14$-$273536.0} & Y & Y & Y & N & Y1 & \omit \\ %g3.0-2.5 & 4740\\
  180403.38$-$333153.0 & N & N & Y & N & Y1 & Y\\ %g357-5.5 & 5387\\
  %180433.70$-$320452.5 & Y & \omit & Y & N & \omit & N\\ %g0.0-5.5 & 21770\\
  180631.55$-$292137.0 & Y & N & Y & N & \omit & \omit \\ %g3.0-5.5 & 16619\\
  {\it 181137.21$-$320650.9} & Y & N & Y & N & \omit & \omit \\ %g0.0-5.5 & 14643\\
  % Now thought to be a star
  %181345.72$-$331738.6 & Y & Y & Y & N & \omit & N\\ %g0.0-8.5 & 17682\\
  181521.88$-$325302.1 & N & N & Y & N & Y1 & N\\ %g0.0-8.5 & 12510\\
  {\it 181634.34$-$323412.4} & Y & N & Y & N & \omit & \omit \\ %g0.0-8.5 & 9003\\
  {\it 181831.06$-$325418.2} & Y & N & Y & N & \omit & \omit \\ %g0.0-8.5 & 9917\\
  183449.11$-$350525.8 & N & N & Y & N & N & N\\ %g0.0-11.5 & 16490\\
  183501.56$-$341820.1 & Y & N & Y & N & Y2& \omit \\ %g0.0-11.5 & 10505\\
  {\it 184154.78$-$360749.2} & N & Y & Y & N & \omit & N\\ %g0.0-14.5 & 23097\\
  185019.66$-$362912.8 & Y & N & Y & N & Y1 & N\\ %g0.0-14.5 & 19904\\
%Now thought to be a star
%185226.04$-$373037.8 & Y & Y & Y & N & \omit & N \\ %X,O,IR: L\,489-43 (odd) %g0.0$-$17.5 25349\\
  190625.66$-$370349.5 & N & N & Y & N & \omit & \omit  \\% Gam\,CrA A (odd) g0.0$-$17.5 12281\\
  190738.78$-$370855.9 & Y & N & Y & N & Y2 & \omit \\ %g0.0-17.5 & 12238\\
\hline
  {\it 170154.67$-$225008.3} & Y & Y & N & Y & \omit & N\\%& g0.0+11.5 10516\\
  {\it 171837.15$-$233510.5} & N & Y & N & Y & N & \omit  \\%& g0.0+8.5 939\\
  {\it 172626.80$-$304809.2} & Y & Y & N & Y & N & N\\%& g357+2.5 13725\\
  {\it 173121.50$-$250235.7} & Y & Y & N & Y & Y1 & \omit\\%&  g3.0+5.5 14570\\
  {\it 174408.71$-$242453.4} & Y & Y & Y & Y & Y1 & \omit \\%& g3.0+2.5 1311\\
  {\it 181822.84$-$324416.0} & Y & Y & N & Y & \omit & \omit \\%& g0.0$-$8.5 8790\\
  {\it 182146.77$-$324727.8} & N & Y & N & Y & \omit & \omit \\%& g0.0$-$8.5 6574\\
% duplicate
  {\it 182417.96$-$323805.4} & Y & Y & Y & Y & \omit & \omit \\%& g0.0$-$8.5 3995\\
  {\it 182731.72$-$325238.9} & Y & Y & N & Y & \omit & \omit \\%& g0.0$-$8.5 3455\\
  {\it 183736.28$-$335013.8} & Y & Y & N & Y & \omit & \omit \\%& g0.0$-$11.5 4028\\
\hline  
  172734.51$-$260957.1 & Y & Y & Y & N & \omit & \omit\\%& g0.0+5.5 10946\\
  {\it 173114.95$-$273532.4} & Y & Y & Y & N & \omit & N\\%& g0.0+2.5 12921\\
  {\it 180221.53$-$325438.0} & Y & Y & Y & N & Y1 & N\\%& g357.0-5.5 1740\\
  {\it 182417.96$-$323805.4} & Y & Y & Y & Y & \omit & \omit \\%& g0.0-8.5 3995\\
  {\it 185120.86$-$354619.5} & Y & N & Y & N & \omit & \omit \\%& g0.0-14.5 12466\\
\enddata
\tablecomments{a. Possibly PSR\,J1753$-$28 (see \S\ref{sec:highenergy})}
\end{deluxetable}

\subsection{Pulsation Searches}\label{sec:pulse}

% Tsky (408/1.4) for the three are 113.5/5.4, 158.0/6.7, 146.8/6.3 all in Kelvin.

While a case can be made that a particular source is a compelling pulsar candidate using imaging and spectra at X-ray, optical and radio wavelengths \citep{2023MNRAS.524.3020K,2024MNRAS.528.5730Z}, the final arbiter is the detection of pulsations. As a pilot project we observed three of the circularly-polarized sources
identified in the G0.0+5.5 mosaic pointing (J172018.40$-$265207.8, J172752.07$-$272724.6 and J172820.26$-$260816.6), using the Ultra-Wide Low-band (UWL; 704-4032 MHz) receiver of the Parkes 64-m radio telescope \citep[Murriyang;][]{2020PASA...37...12H}. The properties of these sources are listed in Table \ref{tab:circ}. For each source, we conducted observations using the UWL for 72 minutes, with a time resolution of 64 microseconds and a frequency resolution of 0.250 MHz, corresponding to 13,312 frequency channels across the 3328 MHz UWL bandwidth. Each observation generated approximately 200 GB of data. We conducted a pulsar search in three different ways using a GPU-accelerated search code, PEASOUP \citep{2019MNRAS.483.3673M}:

\begin{enumerate}
    \item The maximum line-of-sight dispersion measure (DM) for these sources, based on different electron density models, ranges between 250-400 pc cm$^{-3}$ \citep{2002astro.ph..7156C,2017ApJ...835...29Y}. However, we searched over a broader DM range from 2 to 1500 pc cm$^{-3}$. We first divided each UWL observation into four segments of 832 MHz bandwidth each, namely 704-1536 MHz (4345 trial DMs), 1536-2368 MHz (2809 trial DMs), 2368-3200 MHz (1570 trial DMs), and 3200-4032 MHz (853 trial DMs). We then conducted a standard FFT search separately for each segment. Candidates with a spectral SNR above 6 were folded using \texttt{dspsr} and diagnostic plots were obtained using PSRCHIVE's tool \texttt{pdmp} \citep{2011PASA...28....1V,2012AR&T....9..237V}. After folding, candidates with a folded SNR above 8.5 were selected for visual inspection. However, we did not find any significant candidate resembling a real pulsar.

\item We also conducted an acceleration search for the same dataset. The acceleration range searched was $-$30 ms$^{-2}$ to +30 ms$^{-2}$ (which is consistent with the maximum acceleration shown by the majority of the known binary pulsars during 1-hr integration), and the DM range was 2-500 pc cm$^{-3}$. We folded all candidates with a spectral S/N above 7 and visually inspected each with folded S/N above 8.5, but this search also did not yield any significant candidates.

\item Furthermore, we divided the UWL observations into smaller segments, each with a bandwidth of 416 MHz, and conducted an acceleration search similar to that mentioned in (2). This search also did not detect any candidates.

\end{enumerate}

\section{Results}\label{sec:results}

\subsection{Pulsar Candidates}\label{sec:finalcandi}

We have identified a sample of 75 polarized sources in a MeerKAT imaging survey of the bulge of our Galaxy. A total of 60 were found based on their circular polarization (\S\ref{sec:candi}), ten were found from their linear polarization properties (\S\ref{sec:lin-candi}) and five were steep-spectrum polarized sources, detected only in the lower half of the MeerKAT receivers (\S\ref{sec:lowhangingcandi}). We searched for multi-wavelength counterparts to each of these candidates at gamma- and X-ray energies (\S\ref{sec:highenergy}), plus optical and infrared wavelengths (\S\ref{sec:oir}). Based on optical/infrared identifications we found that 28 of the circular polarized sources were bright stars or young stellar objects. Removing these, our final list is shown in the Boolean Table \ref{tab:boolean}. After accounting for duplicates which were detected both in linear and circular polarization, we have a total of 45 sources taken from Tables \ref{tab:circ}-\ref{tab:lowbandcirc}.  

Table \ref{tab:boolean} lists several source properties common to previous image-based searches for pulsars: compactness, spectral steepness, circular polarization, linear polarization, variability, and high energy counterparts. Measures that define how point-like a radio source is, such as the ratio of the total flux over the peak flux density \citep[e.g.,][]{2018MNRAS.475..942F} or the compactness ratio, as defined in \S\ref{sec:catalog} \citep{2017A&A...598A..78I}, have proven useful in identifying pulsar candidates. Despite being unresolved in interferometric images, \cite{2018MNRAS.474.5008D} noted that pulsars may still have compactness less than unity (but usually $>0.9$). This occurs if there are substantial flux density changes within an integration period (scintillation or pulse-to-pulse variations), or the pulsar is located near extended emission (e.g., pulsar wind nebulae). In all cases, even the non-compact sources in Table 
\ref{tab:boolean} have a total flux over the peak flux density ratio $R<$1.5, i.e., they meet the compactness criteria used in \cite{2019ApJ...876...20H} and elsewhere. For spectral steepness we have assigned ``Y" to those with $\alpha<-1.5$ (see \S\ref{sec:lin-candi}), the circular and linear polarization criteria is defined in \S\ref{sec:candi}-\ref{sec:lowhangingcandi} and variability is defined in \S\ref{sec:var}. Finally, we identify all candidates with high energy counterparts (\S\ref{sec:highenergy}). Unlike the eRosita catalog, which covers the entire 4$^{th}$ Galactic quadrant, sky coverage of the \emph{Chandra} catalog is patchy enough that sources in the first Galactic quadrant often lack sufficient data to rule out X-ray emission. 

Of these 45 polarized sources, there are two sources (J174408.71$-$242453.4 and J182417.96$-$323805.4) that are notable for being compact, steep spectrum {\it and} are detected with both strong circular and linear polarization ($>$10\%). Pulsars are the only known compact source population with steep radio spectra to exhibit strong polarization in both forms. These would be high priority candidates for follow-up pulsation searches. Approximately half of our candidates (22) in Table \ref{tab:boolean} are compact, steep spectrum and have linear or circular polarization. With a few exceptions, which we discuss below, the bulk of these sources also appear to be attractive candidates for pulsation follow-up. 

\begin{figure}
%\epsscale{0.8}
% Note range is -4.4 to 2.2, 0 to 110%. Font size=15. Candidates include both circ and linear. Otherwise all other sources are just circular. Grey dots have form=transparent.
\plottwo{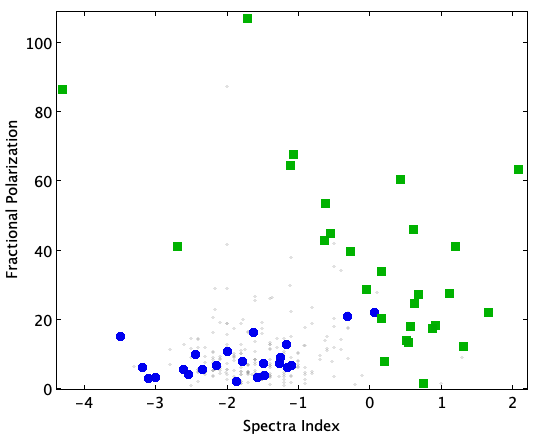}{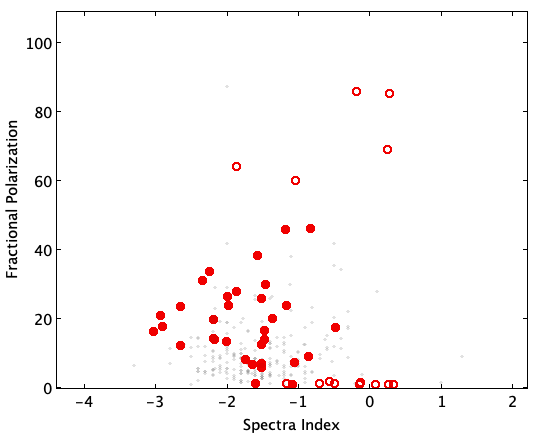}
\caption{(Left) Circular polarization and spectral properties of known pulsars (blue circles) and stars (green squares) identified in Tables \ref{tab:known} and \ref{tab:circ}. Grey points are taken from the compilation of circular polarization and spectral index values of \cite{2018MNRAS.474.4629J} and \cite{2018MNRAS.473.4436J}. (Right) Polarization (circular and linear) versus Spectral Index for pulsar candidates (red) identified in Table \ref{tab:boolean}. Solid circle (open circle) shapes are candidates that lie inside (outside) the known pulsar distribution. The light grey points are the same as the adjacent plot.}\label{fig:specpol}
\end{figure}

Five polarized sources (including J174408.71$-$242453.4) are compact, steep spectrum {\it and} exhibit strong in-band variability (Y1) (see \S\ref{sec:var}). These warrant special attention in pulsation follow-up. The remaining variable sources are flat spectrum and are likely stellar sources. The three high energy counterpart associations are less instructive than the other indicators. In \S\ref{sec:highenergy} we suggest that one source (J175306$-$285126.2) may be a known pulsar, while the remaining two have flat spectral indices similar to the YSOs identified in \S\ref{sec:oir}.

Despite our best efforts at using these multiple selection criteria to select pulsars, there likely remain other source populations among our 45 polarized sources. To illustrate this, we plot the spectral index versus the absolute value of the circular polarization in Fig.\,\ref{fig:specpol}. On the left are the known pulsars (blue) from Table \ref{tab:known} together with the stars (green) identified from \S\ref{sec:oir}. The light grey points are circular polarization and spectral index values taken from the compilations of \cite{2018MNRAS.474.4629J} and \cite{2018MNRAS.473.4436J}, respectively.  The known bulge pulsars trace the spectral index and circular polarization of this larger pulsar sample, except that the larger sample extends to less steep spectral indices and larger polarization ($\leq40$\%). It is clear that the majority of stars occupy different phase space in this diagram. They are distributed over nearly the full range of circular polarization, and most have distinctively flat or inverted spectral indices.

In the right hand side of Fig.\,\ref{fig:specpol} we plot the remaining 45 polarized sources (red) from Table \ref{tab:boolean}. Compared to the larger sample of pulsars (light grey), there is an excess of flat-spectrum candidates in the 1-2\% range and a long tail of strongly polarized sources ($>50$\%). We discussed the excess 
of bright sources in the 1-2\% range in \S\ref{sec:candi}. With the exception of J180221.53$-$325438.0 with $\alpha=-1.61$, we suspect the remaining are extragalactic sources with weak circular polarization or beam squint artifacts.

 The long tail of strongly polarized sources ($>40$\%) in Fig. \ref{fig:specpol} is likely real and not a selection effect or instrumental artifact. Unless there is a pulsar emission geometry that results in a high percentage polarization while making pulsations difficult to detect (i.e. aligned rotators), the bulk of this sample is likely stellar in origin. Many radio stars at this frequency exhibit high fractional circular polarization \citep{2021MNRAS.502.5438P,2024MNRAS.529.1258P}, with properties similar to the stars identified in Table \ref{tab:circ} and Fig. \ref{fig:specpol}. What is more rare are stellar radio stars whose spectral indices are comparable to those of pulsars and suggests a coherent process such as electron cyclotron maser emission \citep{2002ARA&A..40..217G}. Improved astrometry of these steep spectrum, highly polarized candidates will help in identifying multi-wavelength counterparts to show if they are similar to the peculiar J180526.44$-$292954.6 (\S\ref{sec:oir}), or remain interesting pulsar candidates. If we confine the source selection to just those candidates with fractional polarization between 1-45\%, and a negative spectral index, consistent with known pulsar properties, we are left with 30 strong pulsar candidates. The J2000 names of these candidates are highlighted in italics in Table \ref{tab:boolean}.  We note that the excess of positive V sources first discussed in \S\ref{sec:known}, persists in each of these sub-samples. In addition, for the non-pulsar polulations (e.g., YSOs) a positively circularly polarized source is just as likely to have a multi-wavelength counterpart as a negatively polarized source. The implications of this result is discussed more in \S\ref{sec:known}. 
 
To summarize, of the 45 candidates in Table \ref{tab:boolean} that were initially identified on the basis of multiple selection criteria (compactness, spectral index, polarization, variability and multi-wavelength counterparts), we find 30 whose properties closely follow those of the known pulsar population, and are thus good candidates for pulsation follow-up. 

\subsection{Summary of the Candidate Selection Process}

Since this search has been a multi-step process, it is useful to summarized the major steps that have lead to a final list of promising pulsar candidates in section~\ref{sec:finalcandi}. We began with a set of 20 full Stokes images of a 173 deg$^2$ vertical strip of the Galactic bulge. Within this region we identified 387,875 unique radio sources with Stokes I flux density. For each of these sources we measured values (or limits) of Stokes Q, U and V and in-band spectra indices.

To test the efficacy of our methods, we used this catalog to generate a list of the known pulsars that were present in the source catalog. We found polarized emission (linear and circular) from 39 of the 81 known pulsars that lie withing this survey area (Table \ref{tab:known}).
 
As a first step in the selection process, we identified 353 circularly polarized sources above 5$\sigma$. Bright Stokes I sources with $\vert{\rm V/I}\vert<1$\% were eliminated from this sample as likely beam squint artifacts, leaving 100 circularly polarized sources. From visual inspection of the images we identified another 38 likely image artifacts (e.g., sidelobe contamination). After removing the known pulsars we were left with 60 circularly polarized sources (Table \ref{tab:circ}). 
 
A similar method was followed to identify linearly polarized sources but owing to high source density in the linear polarized images, there was a much great incidence of artifacts. As a result, additional selection criteria were required (compactness, spectral index and fractional polarization) to keep the false positives to a manageable level. After removing likely artifacts the final sample has 10 linearly polarized sources (Table \ref{tab:linear}). As a final method, we confided our search to polarized sources in the low half of the MeerKAT band to catch fainter, steep spectrum sources missed in our full-band searches. This resulted in five additional circularly polarized sources (Table \ref{tab:lowbandcirc}). In total these searches resulted in 75 polarized sources (Tables \ref{tab:circ}-\ref{tab:lowbandcirc}). 
 
The next step in the selection process was to search for multi-wavelength counterparts of our polarized sources. We identified 28 bright stellar sources (YSOs, M stars, pulsating variable, etc.) with high confidence. After removing these sources and accounting for two duplicate sources which were both circularly and linearly polarized, we are left with a sample of 45 polarized sources. While all of these are viable pulsar candidates, it is clear from their aggregate properties there still remain non-pulsars or pulsar outliers in this sample. To further refine this polarized sample we examine other known pulsar properties including compactness, spectrum, variability and high energy counterparts.  A final list of 30 promising candidates (italic source names in Table \ref{tab:boolean}) is obtained by requiring that these source properties are in line with known pulsars. 

\subsection{The Missing Pulsations}\label{sec:missing}

%\textcolor{red}{WORK IN PROGRESS.}

There are 81 pulsars that have been previously discovered toward these bulge mosaic fields (\S\ref{sec:known}). In this work we have identified 30 additional pulsar candidates. The immediate question that needs to be answered is why have previous pulsation searches, including \S\ref{sec:pulse}, not detected pulsations toward these sources? The myriad of reasons for non-detection of pulsations have been previously discussed in the literature and broadly fall into four categories: instrumental, algorithmic, environmental and intrinsic. The challenges and strategies for finding pulsations toward the bulge and the GC have been discussed extensively \citep{2015ApJ...805..172M,2016ApJ...827..143C,2019ApJ...876...20H,2021MNRAS.507.3888H}.

Instrumental sensitivity appears to be part of the explanation. The deepest pulsation surveys at 1.4 GHz that cover the MeerKAT bulge mosaic images are the Parkes High Time Resolution Universe survey \citep[HTRU;][]{2010MNRAS.409..619K} and the Parkes Multi-beam Pulsar Survey \citep[PMPS;][]{2001MNRAS.328...17M}. The majority of the known 81 pulsars were discovered by the PMPS and HTRU. Thanks to improvements in search algorithms, the re-processing of these datasets has lead to additional new pulsar discoveries close to the continuum sensitivity limits \citep{2020MNRAS.493.1063C,2023MNRAS.522.1071S}

The PMPS surveyed $\vert{b}\vert<5^\circ$, while HTRU  had three parts; HTRU-low with $\vert{b}\vert<3.5^\circ$, HTRU-mid with  $\vert{b}\vert<15^\circ$ and HTRU-high for all other Galactic latitudes. The integration time at the low, mid and high latitudes was 4300 s (twice PMPS), 540 s and 270 s, respectively. HTRU had smaller channel bandwidths than PMPS, making it less limited by dispersion and scattering, thus increasing its sensitivity to distant MSPs. The mean limiting flux density of the HTRU has a strong latitude dependence in part because of the decreasing integration times from low, to mid to high, but also because of the variation in sky temperature \citep{2014PASA...31....7C,2015MNRAS.451.4311R}. Taken together, we estimate that the HTRU-high and HTRU-mid was 3.5x and 2.6x less sensitive than the HTRU-low. 

\begin{figure}[htb!]
\includegraphics[width=10cm]{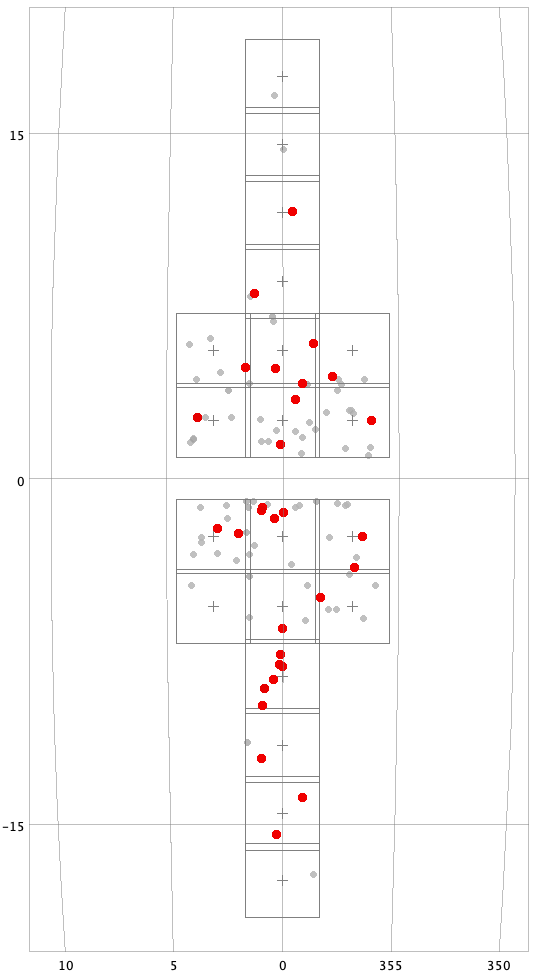}
\centering
\caption{Galactic distribution of the known pulsars in the mosaic fields (light grey crosses) versus our final 30 pulsar candidates (red circles).}\label{fig:goodsky}
\end{figure}

In contrast, the rms noise for the MeerKAT bulge mosaic images do not show a strong Galactic latitude dependence. The noise is relatively constant with a mean of 14 $\mu$Jy bm$^{-1}$ (see Table \ref{tab:point}), varying by $\sim$30\% owing to changes in T$_{sky}$ and varying degrees of RFI. Using the radiometer equation and the instrumental parameters of the HTRU survey for the outer beams of the 13-beam receiver \citep{2010MNRAS.409..619K,2020MNRAS.493.1063C}  we expect that the limiting flux density of the MeerKat polarized pulsar candidates to be very similar to pulsars previously seen by HTRU-low. This indeed appears to be the case. We find that the limiting flux density for the both the HTRU detections and the circularly polarized candidates in Tables \ref{tab:circ} and \ref{tab:lowbandcirc} are both approximately 0.1 mJy. Since our pulsar candidates are distributed over $\vert{{b}\vert}<20^\circ$ (see Fig.\,\ref{fig:goodsky}), the sharp decrease in the HTRU sensitivity with latitude likely explains why the fainter pulsar candidates were missed in previous pulsation searches. 

Sensitivity cannot explain why the HTRU previously did not detect pulsations from the brighter candidates. such as J174408.71$-$242453.4; a 1.2 mJy source that is compact, steep spectrum and has both significant linear {\it and} circular polarization. The sky distribution of our sample, compared to existing pulsation surveys, offers some clues (Fig.\,\ref{fig:goodsky}). There are two groupings of candidates in the mosaic fields; a group of seven distributed along a narrow band of longitude primarily in the G0.0$-8.5$ mosaic, and a less well-defined group near b=5$^\circ$.  While this clustering could be random, we note that they occur in the direction of ``voids'' wherein there are no known pulsars. We suggest that anomalous scattering might be responsible for one or more of these voids. An excellent case in point is the recent discovery of the highly scattered pulsar PSR\,J1032$-$5804 from a imaging survey of polarized sources \citep{2024ApJ...961..175W}. 

If sensitivity and interstellar scattering are the main culprits then deeper searches and/or searches above L band should be successful in detecting pulsations. Neither of these appears to explain the lack of pulsations from the three candidates observed in \S\ref{sec:pulse}. The search of these mid-latitude candidates was conducted with the Parkes telescope using the same integration time as HTRU-low, over a much wider frequency range (704-4032 MHz), and using state-of-the-art search algorithms. Using the instrumental parameters from \S\ref{sec:pulse} and T$_{sky}$ at the source positions, we used the radiometer equation to calculate the rms noise\footnote{\url{https://www.parkes.atnf.csiro.au/cgi-bin/utilities/pks_sens.cgi}}. For example, at 1400 MHz, T$_{sky}$=6.7 K, and a bandwidth of 1024 MHz, and a S/N threshold of 8.5, the sensitivity is 0.1 mJy$\times\sqrt{w_t/({\rm{P}}-w_t)}$, where P is the pulsar period and $w_t$ is the quadrature sum of the intrinsic pulsar width, temporal scattering, dispersive smearing and the sampler timescale \citep{2023MNRAS.522.1071S}. We then modeled an isolated MSP with P$\sim$5 ms with a 5-10\% pulse width at a distance of 8300 pc, estimating its dispersion measure and the degree of scattering using two Galactic electron density models \citep{2002astro.ph..7156C,2017ApJ...835...29Y}. We find that pulsations should have been expected for this putative bulge MSP at high significance over much of the frequency range of the UWL receiver. Perhaps some of the candidates are rare and more exotic systems such as tight binaries or pulsars with broad emission profiles \citep{1995ApJ...455L..55N,2019ApJ...884...96K}, or, as has been suggested, represent a new galactic population with properties like pulsars but lacking pulsations \citep{2021MNRAS.507.3888H}.

\section{Discussion and Conclusions}\label{sec:discuss}

We have carried out an analysis of a sensitive MeerKAT survey of the Galactic bulge and GC region at L band (856-1712 MHz) in all four Stokes  I, Q, U and V images. The survey covers 173 deg$^2$ in 20 partially overlapping mosaic pointings, each 3.125$^\circ$×3.125$^\circ$. Our primary science driver has been to identify pulsar candidates, with a long-term goal of explaining the origin of the mysterious gamma-ray emission identified by the \emph{Fermi} Gamma-Ray Telescope.

Previous image-based candidate searches have been plagued by contaminants (variables, extragalactic sources, etc.) \citep[e.g.,][]{2000A&AS..143..303D,2000ApJ...529..859K,
2018MNRAS.474.5008D,2018MNRAS.475..942F,2018ApJ...864...16M,2023ApJ...943...51B}. This current work is a substantial improvement over past image-based pulsar searches in several dimensions: sensitivity, improved angular resolution/astrometry, polarization,  in-band spectra, and the use of multiple selection criteria. The rms noise levels of these mosaic images is 12-17 $\mu$Jy ba$^{-1}$ (see Table \ref{tab:point}). When past continuum surveys such as the NRAO VLA Sky Survey \citep[NVSS;][]{1998AJ....115.1693C}, the TIFR GMRT Sky Survey \citep[TGSS;][]{2017A&A...598A..78I}, and the Galactic and Extragalactic All-sky Murchison  Widefield Array (MWA) survey \citep[GLEAM;][]{2017MNRAS.464.1146H} are scaled to L band, using a mean pulsar spectral index, this MeerKAT bulge survey is nearly a two orders of magnitude increase in sensitivity. Using the MeerKAT Absorption Line Survey \citep[MALS;][]{2024ApJS..270...33D}, Himes et al. (2024, in prep.) showed that these earlier surveys were sensitive to between 7-21\% of the {\it known} pulsar population, while for MALS the fraction was 80\%. The same conclusion applies to this MeerKAT bulge survey since it is approximately a factor of two deeper than MALS. More importantly, for the first time the continuum sensitivity has reached a level that is capable at detecting the putative population of MSPs in the Galactic bulge, which can be responsible for the {\it Fermi} excess and whose properties have been described by \cite{2016ApJ...827..143C}.

The next generation of all-sky surveys capable of imaging the GC and bulge will have sensitivities comparable to MeerKAT. One area that MeerKAT retains an advantage over ASKAP \citep{2021PASA...38...54M,2022MNRAS.516.5972W} and MWA \citep{2022PASA...39...35H} is in angular resolution. Higher resolution can be useful for identifying false positives such as extended, steep-spectrum high redshift galaxies \citep{2008A&ARv..15...67M}, but for a crowded region such as the bulge, accurate astrometry is essential to identify multi-wavelength counterparts (\S\ref{sec:count}). 

Another improved capability of these next generation facilities is the ability to provide full Stokes images. At these sensitivity levels, just using criteria such as compactness and spectral index \citep[e.g.,][]{2000ApJ...529..859K,
2018ApJ...864...16M,2018MNRAS.474.5008D} produces too many candidates. For example, if we select all compact and steep spectral index sources in the G0.0+5.5 mosaic using the criteria in \S\ref{sec:lin-candi} we find nearly 900 candidates, or nearly 4\% of the cataloged radio sources for a source density of 90 deg$^{-2}$. Including a polarization criteria to this sample (\S\ref{sec:candi}) sharply reduces this source density. For this bulge survey region, if we include the known pulsars in Table \ref{tab:known} and the circularly polarized candidates in Tables \ref{tab:circ} and \ref{tab:lowbandcirc}, the source density drops to a manageable 0.5 deg$^{-2}$. 

The large fractional bandwidths allow for the measurement of the spectral index at one time, reducing the rate of false positives. Past searches determined the spectral index from two or more surveys at different frequencies but often taken at very different epochs \citep[e.g.,][]{2016ApJ...829..119F,2024ApJ...969...30M}. \cite{2018MNRAS.474.5008D} argued that the tail of the spectral index distribution at the steep end could be contaminated by variable extragalactic sources.  To our knowledge, this is the first large-scale study to identify pulsar candidates using both spectral indices and polarization. As we showed in \S\ref{sec:finalcandi}, the polarization/spectral index space (Fig. \ref{fig:specpol}) can be a powerful tool for distinguishing different source populations. As an added benefit this method should allow us to identify promising pulsar candidates without restricting the search to the steepest spectral indices. Finally, we note that this work has followed the recent trend of using multiple selection criteria (variability, compactness, steep spectrum, polarization and multi-wavelength counterparts) to bolster claims of pulsar candidates \citep{2023PASA...40....3S}. 

After applying these selection criteria, we identify 45 pulsar candidates (\S\ref{sec:finalcandi}). We noted that there likely remained some false positives in the sample, owing in part to instrumental artifacts at low fractional polarization (1-2\%) and with the difficulty in identifying optical/infrared counterparts in these crowded GC fields with our current astrometric accuracy (\S\ref{sec:properties}). Focusing on just those which have observational properties similar to the known pulsars we find 30 strong candidates for pulsation follow-up. We argue that the absence of pulsations from these candidates from previous surveys is mainly due to a combination of low sensitivity and interstellar scattering, but other factors must be involved.

The challenges facing pulsation searches in the Galactic bulge is demonstrated by the recent discovery of the first MSP close to the Galactic center. X-ray and radio images showed a compact source embedded within a faint filament west of the well-known non-thermal source G359.1$-$0.2 (the "Snake"), but it took a {\it targeted} search to find PSR\,J1744$-$2946 with a period P=8.4 ms \citep{2024ApJ...967L..16L}. The pulsar was visible on MeerKAT images at 1.4 GHz but its pulses were scatter broadened, while the same pulses were bright at 2.1 GHz (0.43 mJy). This explains why it was missed by the HTRU survey and highlights the importance of wide-bandwidth searches. If PSR\,J1744$-$2946 is located within the Galactic bulge, it may lie on the high-luminosity tail of the putative bulge pulsar population discussed by \citet{2016ApJ...827..143C}. This conclusion also applies to the pulsar candidates identified here because our polarization criteria limits detection to $\gtrsim$0.1 mJy. In order to exploit the full power of these sensitive Stokes I MeerKAT observations in image-based pulsar searches, other multi-wavelength approaches based of cross-correlation techniques may be needed \citep{Berteaud:2023xjp,Berteaud:2024MeerKAT}.

\begin{acknowledgments}

The MeerKAT telescope is operated by the South African Radio Astronomy Observatory, which is a facility of the National Research Foundation, an agency of the Department of Science and Innovation.
Murriyang, the Parkes radio telescope, is part of the Australia Telescope National Facility which is funded by the Australian Government for operation as a National Facility managed by CSIRO. The National Radio Astronomy Observatory is a facility of the National Science Foundation operated under cooperative agreement by Associated Universities, Inc.
This work has made use of data from the European Space Agency (ESA) mission
{\it Gaia} (\url{https://www.cosmos.esa.int/gaia}), processed by the {\it Gaia} Data Processing and Analysis Consortium (DPAC,
\url{https://www.cosmos.esa.int/web/gaia/dpac/consortium}). Funding for the DPAC has been provided by national institutions, in particular the institutions participating in the {\it Gaia} Multilateral Agreement. RS and DK are supported by NSF grant AST-1816904. 
Parts of the results in this work make use of the colormaps in the CMasher package \citep{cmasher}. Basic research at NRL is funded by 6.1 Base programs. This project was supported in part by an appointment to the NRC Research Associateship Program at the US Naval Research Laboratory, administered by the Fellowships Office of the National Academies of Sciences, Engineering, and Medicine.

\end{acknowledgments}

\vspace{5mm}
\facilities{MeerKAT, Parkes}


\software{PyBDSF \citep{2015ascl.soft02007M}, astropy \citep{2013A&A...558A..33A,2018AJ....156..123A},  
TOPCAT \citep{2005ASPC..347...29T,2011ascl.soft01010T},
%          Source Extractor \citep{1996A&AS..117..393B}
          }

\bibliography{polarcandi}{}
\bibliographystyle{aasjournal}



\end{document}